\documentclass[twocolumn]{aastex631}
\usepackage{outlines}

\shorttitle{Broad line measurements in nearby megamasers}
\graphicspath{{./}{figures/}}

\begin{document}

\title{Spectropolarimetric measurements of hidden broad lines in nearby megamaser galaxies: a lack of clear evidence for a correlation between black hole masses and virial products\footnote{based on observations made with the Southern African Large Telescope (SALT)}}

% \correspondingauthor{}
% \email{}

\author[0000-0001-8840-2538]{Nora B. Linzer}
\affiliation{Department of Astrophysical Sciences, Princeton University, 4 Ivy Lane, Princeton, NJ 08854, USA}

\author[0000-0003-4700-663X]{Andy D. Goulding}
\affiliation{Department of Astrophysical Sciences, Princeton University, 4 Ivy Lane, Princeton, NJ 08854, USA}

\author[0000-0002-5612-3427]{Jenny E. Greene}
\affiliation{Department of Astrophysical Sciences, Princeton University, 4 Ivy Lane, Princeton, NJ 08854, USA}

\author[0000-0003-1468-9526]{Ryan C. Hickox}
\affiliation{Department of Physics and Astronomy, Dartmouth College, 6127 Wilder Laboratory, Hanover, NH 03755, USA}

\begin{abstract}
High-accuracy black hole (BH) masses require excellent spatial resolution that is only achievable for galaxies within $\sim$100 Mpc using present-day technology. At larger distances, BH masses are often estimated with single-epoch scaling relations for active galactic nuclei. This method requires only luminosity and the velocity dispersion of the broad line region (BLR) to calculate a virial product, and an additional virial factor, $f$, to determine BH mass. The accuracy of these single-epoch masses, however, is unknown, and there are few empirical constraints on the variance of $f$ between objects. We attempt to calibrate single-epoch BH masses using spectropolarimetric measurements of nine megamaser galaxies from which we measure the velocity distribution of the BLR. We do not find strong evidence for a correlation between the virial products used for single-epoch masses and dynamical mass, neither for the megamaser sample alone or when combined with dynamical masses from reverberation mapping modeling. Furthermore, we find evidence that the virial parameter $f$ varies between objects, but we do not find strong evidence for a correlation with other observable parameters such as luminosity or broad line width. Although we cannot definitively rule out the existence of any correlation between dynamical mass and virial product, we find tension between allowed $f$ values for masers and those widely used in the literature. We conclude that the single-epoch method requires further investigation if it is to be used successfully to infer BH masses.
\end{abstract}

\keywords{}

\section{Introduction}

Water megamasers are extremely luminous sources of 22 GHz radiation generated by the amplification of microwave signals through stimulated emission \citep[for a review, see][]{lo2005mega}. They can be found within a few parsecs of active galactic nuclei (AGN) and may be used to probe the kinematics of this inner region \citep[e.g.][]{greenhill1996vlbi, trotter1998water, peck2003flaring}. Some special disk systems, such as NGC 4258, have masers which trace the ridge-line of an edge on disk, allowing for precise measurements of the disk dynamics \citep{herrnstein1999geometric}. Measurement of the acceleration of systemic features provide an independent evaluation of H$_0$ \citep[e.g.][]{kuo2013megamaser, kuo2015megamaser, reid2013megamaser, gao2016megamaser, pesce2020megamaser}. The rotation axis of the maser disk is also observed to align with the jet axis when jets are detected, suggesting the maser disk can be used to understand the geometry of the accretion disk \citep[e.g.][]{greene2013using, kamali2019accretion}.

In addition, the Keplerian rotation of the sub-parsec scale accretion disk traces the black hole (BH) mass with high precision and accuracy \citep[e.g.][]{miyoshi1995evidence}. Currently, there are only $\sim100$ BHs with masses determined by dynamical tracers such as masers, stars, or gas including both active and non-active galaxies \citep[e.g.][]{kormendy2013coevolution, mcconnell2013revisiting, saglia2016sinfoni}, as this method requires that the BH sphere of influence be spatially resolved. Therefore, with current adaptive optics we are limited to objects within $\sim 100$ Mpc for dynamical estimates, except for the most massive BHs. No other BH has a mass measured with the same precision as that in the Galactic Center, but the masers offer the most precise measurement after that \citep[e.g.][]{maoz1998dynamical, kuo2010megamaser}. Beyond the available dynamical masses, all other BH masses have been estimated using only indirect tracers, often involving accretion \citep[e.g.][]{shen2013mass}. We will use the high accuracy and precision of the maser masses to test the fidelity of other methods of BH mass measurements, specifically single-epoch scaling relations in AGN.

In the absence of available maser measurements, the most accurate method for determining BH masses using emission from AGN is reverberation mapping (RM). This approach uses broad line region (BLR) gas that is not spatially resolved as a dynamical tracer to measure velocity from line widths. While it is not possible to resolve the gravitational sphere of influence in most of these AGN, it is possible to determine a size scale for the broad line region by measuring the time lag between variations in the AGN continuum to those in the emission lines of the BLR \citep[e.g.][]{blandford1982reverberation, peterson1993reverberation}. This delay provides an estimate of the light-travel time across the BLR, from which the average radius of the BLR can be determined. The BLR radius (R), in combination with the gas velocity measured from the width of the broad emission lines (W) and gravitational constant (G), can then be used to calculate a virial product (Equation \ref{eq:vp}). 

\begin{equation}
\label{eq:vp}
\textrm{virial product} = \frac{W^2 R}{G}
\end{equation}

Under the assumption that the BH dominates the gravity in this region, the virial product is expected to be correlated with the mass of the BH, but there is a virial pre-factor, $f$, required to account for the geometry and dynamics of the disk. The pre-factor is defined such that $f$ multiplied by the virial product gives the virial BH mass (Equation \ref{eq:mass}). 

\begin{equation}
\label{eq:mass}
M_{\textrm{BH}} = f \frac{W^2 R}{G}
\end{equation}

In general, we do not know the shape or kinematics of the BLR, whether it is flattened or round, or whether the gas is inflowing, outflowing or neither. In a growing number of cases, this has been modeled \citep[e.g.][see more discussion below]{pancoast2014modelling, grier2017structure, 2018ApJ...869..137L, 2022ApJ...927...58L, williams2018lick, williams2020space, bentz2021detailed, Villafa_a_2022} but in general, an average value of $f$ has been used.

As well as providing an estimate of the BH mass, RM has been used to calibrate a relationship between the luminosity of the AGN and the BLR size \citep[e.g.][]{bentz2013low}. With this relationship, the BH masses of distant AGN can be estimated through the virial product with only luminosity and a measured line width \citep{vestergaard2002determining}, known as the single-epoch method. Most inferences about the cosmic evolution of BH mass density, and potential evolution in BH-galaxy scaling relations, have relied on these single-epoch virial masses \citep[e.g.][]{laor1998quasar, wandel1999central, mclure2002measuring, vestergaard2006determining, kelly2013demographics, volonteri2016inferences, pensabene2020alma}. Therefore, it is crucial to determine whether the virial product provides an accurate value of BH mass. 

The virial estimate relies on multiple assumptions about the structure of the BLR. Uncertainties in calibrations alone lead to estimates of 0.3-0.5 dex scatter \citep{vestergaard2006determining, shen2013mass}. There are also as-yet unquantified systematic errors that are still a large cause for concern. Primarily, the virial estimate assumes that broad emission lines are virialized \citep[e.g.][]{peterson1999keplerian, peterson2000evidence, onken2002mass, kollatschny2003accretion, bentz2010lick}. Additionally, the gas velocity of the BLR is assumed to be at the radius measured by reverberation mapping, but this is not necessarily the case \citep{krolik2001systematic}. Although gravity is assumed to dominate in the BLR, there is an unknown contribution from radiation pressure \citep[e.g.][]{krolik2001systematic, marconi2008effect, marconi2009observed, netzer2010effect}. This could introduce additional scatter, as well as a luminosity dependence \citep{shen2013mass}, which may also be generated from BLR breathing modes \citep{wang2020sloan}.

The determination of an accurate black hole mass also depends on the choice of $f$, which in turn must be calculated through an independent method. Historically, $f$ has been calibrated by aligning the M-$\sigma_*$ relation of reverberation mapped AGN with that of quiescent galaxies \citep[e.g.][]{onken2004supermassive, collin2006systematic, woo2010lick, graham2011expanded, grier2013stellar, batiste2017recalibration}. Although $f$ is often taken to be a constant, it has been found to vary significantly between objects \citep{yu2019calibration}. An independent method of determining $f$ by modeling the BLR has also found variation in the virial pre-factor \citep[e.g.][]{pancoast2014modelling, grier2017structure, williams2018lick}. This modeling is intensive, and requires densely sampled light curves, so it has only been done for 27 objects so far \citep{Villafa_a_2022}. More comparisons are needed.

Evaluating the accuracy of the virial product as a probe of BH mass requires a comparison to a known, dynamical mass. In general, this is very challenging since luminous AGN are needed for RM, but severely complicate stellar or gas dynamical measurements, as their light swamps that of the stars. Although there are a few objects with both measurements \citep[e.g.][]{davies2006star, onken2007black,den2015measuring}, building up this sample will be slow. We instead use the larger sample of objects with maser dynamical masses. However, the accretion disk is at very high inclination to enable masing, so the BLRs of maser galaxies are all obscured. Therefore, we must use spectropolarimetric measurements to probe the BLR and measure the hidden broad lines so that we may calculate a virial product. Similar work has been done previously for smaller samples of megamaser galaxies \citep{kuo2010megamaser, du2017hidden}. By expanding this sample through additional spectropolarimetric measurements of maser galaxies, we sought to test the hypothesis that a direct correlation exists between virial product and BH mass.

In Section \ref{sec:methods} we describe our spectropolarimetric observations of megamaser galaxies and subsequent data reduction. We present our measurements of broad line widths, along with additional values from the literature, in Section \ref{sec:results}. BH masses determined through the virial product or through RM modeling are given in Section \ref{sec:BHmass}. Section \ref{sec:disc} includes discussion of the virial pre-factor, and implications for the virial mass and BLR structure. We summarize our results in Section \ref{sec:summary} and discuss possible future work.

\section{Methods}
\label{sec:methods}

\subsection{Sample}

We began with the sample of megamaser galaxies with known, dynamical BH masses listed in \cite{kuo2020megamaser}. The dynamical masses were determined through the work of the Megamaser Cosmology Project \citep[MCP;][]{ reid2009megamaser, braatz2010megamaser}. We selected any megamaser disk with a published BH mass, even if double-peaked rather than triple-peaked with more complex kinematics. Uncertainties in the BH mass are adopted from \cite{kuo2020megamaser} or \cite{greene2016megamaser} based on the dynamical modeling papers referenced therein.

Of the 22 megamasers included in the parent sample, we observed nine as described in Section \ref{sec:data}. Among these nine objects, we find evidence of a polarized broad line in three (Section \ref{sec:fit}). In addition to the nine we observed, we include six additional galaxies with measured polarized broad lines in the literature. Our complete sample of objects with broad line widths is described in Section \ref{sec:results}.

\subsection{Data}
\label{sec:data}
Linear spectropolarimetry of nine megamaser galaxies with known disk dynamical BH masses was obtained with the Robert Stobie Spectrograph (RSS) on the Southern African Large Telescope (SALT). See Table \ref{tab:sample} for dates of observation and exposure times. Each of the nine objects was observed on one night, except for NGC 1194 which was observed on two. Exposure time was divided evenly between four waveplate angles, with three observations at each angle. The seeing was approximately 0.6$\arcsec$. Resolution was $R \approx 1065$, and the pixel scale was 0.1267$\arcsec$ per pixel. The spectra were taken in a wavelength range of 4200\AA - 7270\AA.

\begin{deluxetable*}{lcclc}
\tablecaption{\label{tab:sample} Observation Megamaser Sample}
\tablewidth{0pt}
\tablehead{
\colhead{Galaxy} & \colhead{Distance (Mpc)} & \colhead{Ref.} & \colhead{Date Observed} & \colhead{Exposure Time (s)}
}
\decimalcolnumbers
\startdata
IC 2560 & 41.8 & 1 & 2017-05-16  & 1440 \\
Mrk 1029 & 124.0 & 1 & 2017-10-13 & 2280 \\
NGC 1068 & 15.9 & 1 & 2017-08-26 & 1200 \\
NGC 1194 & 53.2 & 2 & 2017-10-11, 2017-10-17 & 1440, 1440 \\
NGC 1320 & 49.1 & 2 & 2017-10-11 & 1440 \\
NGC 2960 & 49.1 & 1 & 2017-05-14 & 1440 \\
NGC 3393 & 49.2 & 1 & 2017-05-22 & 1200 \\
NGC 5495 & 93.1 & 2 & 2017-05-20 & 1440 \\
NGC 5765b & 126.3 & 2 & 2017-05-22 & 2400 \\
\enddata
\tablecomments{Observed galaxy sample. Columns 2-3 give the distance to the megamaser and references. Columns 4-5 provide the date of observation and exposure time. References: (1) \cite{greene2016megamaser}, (2) \cite{kuo2020megamaser} and included references.}
\end{deluxetable*}

\subsection{Reduction}
\label{sec:reduction}

The data were reduced with the polsalt\footnote{https://github.com/saltastro/polsalt} extension to the pysalt\footnote{http://pysalt.salt.ac.za/} \citep{crawford2010pysalt} reduction pipeline with a few minor modifications. Basic reduction steps include overscan subtraction; corrections for gain, crosstalk, and distortion; and cosmic ray cleaning. We modified the wavelength calibration method slightly to ensure that the wavelength was fit over the full pixel domain. After wavelength calibration, individual spectra were extracted by manually selecting the center and width. The O and E spectra for each observation were interpolated to use the same wavelength solution, then combined to calculate the Stokes parameters. 

At this step, the reduction pipeline was modified to account for masked pixels. In the original software, if any of the three observations at a given waveplate angle had a masked pixel at a certain wavelength, the corresponding pixel in the combination would be masked. We altered the reduction such that if only one pixel out of three were to be masked, that pixel would be replaced with the average value of the remaining two pixels.

The Stokes parameters were combined to generate the total intensity, polarization fraction (P), and polarization angle ($\theta$) for each object. The resulting spectra are missing $\sim$50 \AA\ of data between $\sim$5220-5270 \AA\ and $\sim$6260-6310 \AA\ due to the location of the chip gaps. These gaps do not affect the analysis of the broad H$\alpha$ region.

Before fitting the broad H$\alpha$ line (\S \ref{sec:fit}), we performed additional continuum subtraction from the Stokes parameters following the method described in \cite{capetti2021spectropolarimetry}. We estimated the continuum polarization for each object by taking regions of 30-80 \AA\ on either side of the H$\alpha$ line, then performing a constant fit to the values of I, Q, and U between these regions. The continuum fit was then subtracted before the Stokes parameters were combined to find P and $\theta$. The regions used for the background fit were chosen in each object to avoid emission and absorption lines, as well as the chip gaps. This subtraction improved the detection of the polarized broad lines and removed interstellar polarization in the region of interest.

\subsection{Standard Star}

To ensure the observations of the nine sample objects are properly calibrated, we observed a standard polarized star, BD-12 5133. This star has a known polarization fraction of 4.27 $\pm$ 0.02 \% in the $V$-band, and a polarization angle of 145.88 $\pm$ 0.09$^{\circ}$ \citep{CikotaStandard}. We apply the polsalt data reduction to the standard star and measure a polarization fraction of 4.6 $\pm$ 0.2 \% and a polarization angle of 144 $\pm$ 1$^{\circ}$ averaged over the V band (5070 - 5950 \AA). The values of polarization fraction and angle are consistent within 2$\sigma$ so we are assured the reduction is well calibrated, although we discuss a possible calibration issue in Section \ref{sec:spectra}.

\section{Results}
\label{sec:results}

\subsection{Spectra}
\label{sec:spectra}

For each object, we show P and $\theta$ in 25 \AA\ bins as well as the total and polarized intensity for the full observed wavelength spectrum. We also show total and polarized intensity in the H$\alpha$ region after performing continuum subtraction. One example is shown for IC 2560 in Figure \ref{fig:ic2560_spectra}, and the rest in Appendix A.

\begin{figure*}[htb!]
    \plotone{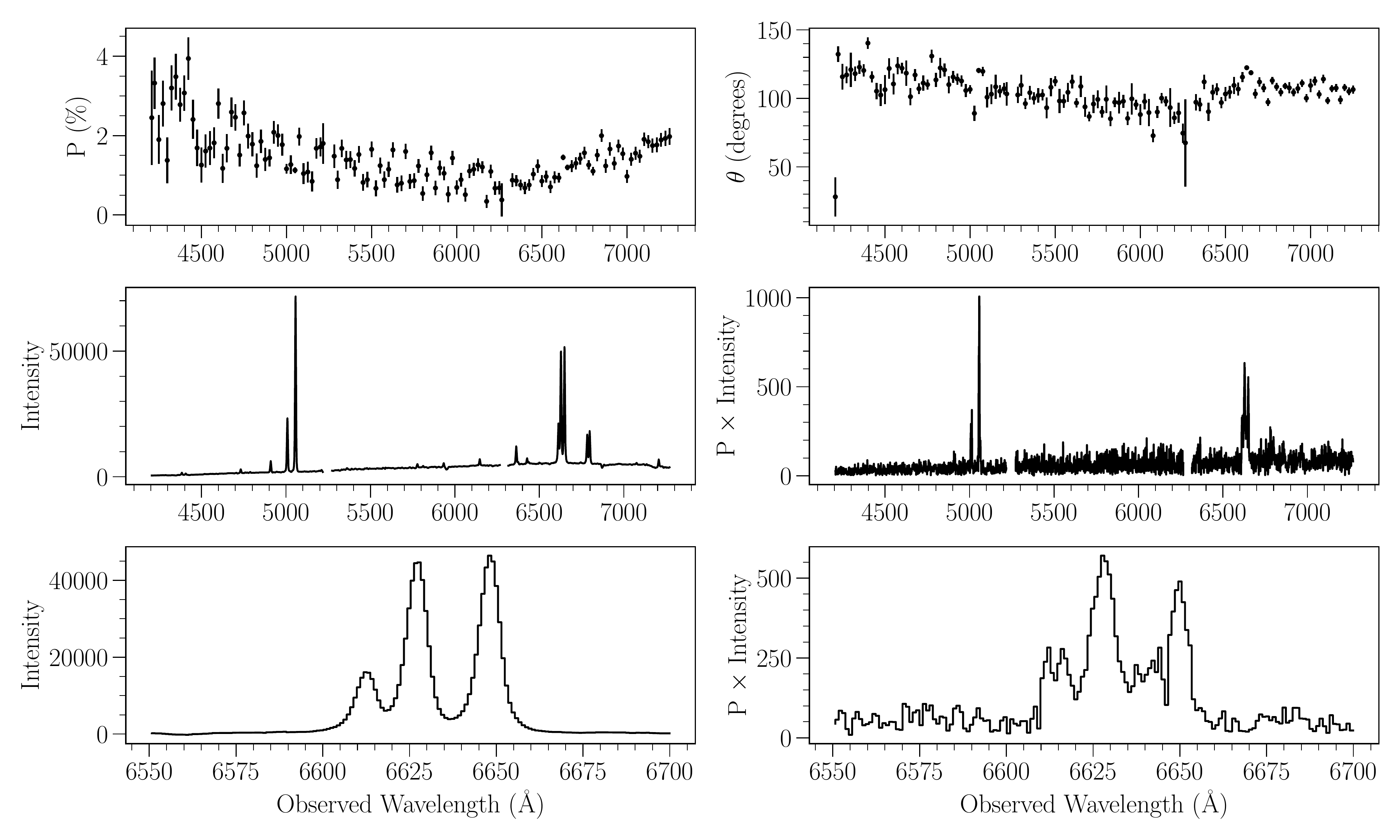}
    \caption{Spectrum of IC 2560. Top row: polarization fraction and polarization angle in 25 \AA\ bins. Middle row: total and polarized intensity in arbitrary units for the full wavelength range. Bottom row: total and polarized intensity in the H$\alpha$ region after continuum subtraction.}
    \label{fig:ic2560_spectra}
\end{figure*}

We find the average $\theta$ and P in the V-band continuum (5800 - 6300 \AA) and across H$\alpha$ (6500 - 6625 \AA) in the rest frame wavelength for each object following \cite{ramos2016upholding}. We also find the signal-to-noise ratio of P in the same regions. These values are given in Table \ref{tab:pol}. We find that $\theta$ is not well determined due to large scatter over short wavelength ranges. Our values have fractional errors of $\sim30\%$. A precise value of $\theta$, however, is not important for our analysis of the virial product.

We find the results of P and $\theta$ for IC 2560 and NGC 3393 to be consistent within errors when compared to the observations in \cite{ramos2016upholding}. Our measurements of NGC 1068 agree with the results presented in \cite{inglis1994spatially} and \cite{young1995near} for observations of the nucleus.

The value of P increases towards the red end of the spectrum for both IC 2560 and NGC 5765b (Figure \ref{fig:ic2560_spectra}). This is most likely an issue with calibration rather than S/N as we do not see the same feature in the other objects in the sample all with similar values of S/N. This might be indicative of a red galaxy continuum present in our polarized spectra. However, as we subtract the total continuum before fitting the broad feature (\S \ref{sec:reduction}), this issue should not affect our measured values.

\begin{deluxetable*}{lcccccc}
\tablecaption{\label{tab:pol} Polarization Angle and Fraction}
\tablewidth{0pt}
\tablehead{
\colhead{Galaxy} & \colhead{$\theta_V$ (deg)} & \colhead{$\theta_{H\alpha}$ (deg)} & \colhead{P$_V$(\%)} & \colhead{P$_{H\alpha}$(\%)} & \colhead{SNR$_V$} & \colhead{SNR$_{H\alpha}$}
}
\decimalcolnumbers
\startdata
        IC 2560 & 100 $\pm$ 40 & 120 $\pm$ 20 & 1.6 $\pm$ 0.9 & 1.3 $\pm$ 0.6 & 4.5 &  6.0 \\
        Mrk 1029 & 0 $\pm$ 40 & 0 $\pm$ 30 & 2 $\pm$ 1 & 1.0 $\pm$ 0.6 & 4.6 &  3.6 \\
        NGC 1068 & 88 $\pm$ 6 & 90 $\pm$ 10 & 2.6 $\pm$ 0.6 & 1 $\pm$ 2 & 70.3 &  364.3 \\
        NGC 1194 & 150 $\pm$ 30 & 150 $\pm$ 10 & 2 $\pm$ 1 & 2.2 $\pm$ 0.7 & 6.4 &  10.4 \\
        NGC 1320 & 110 $\pm$ 30 & 110 $\pm$ 30 & 3 $\pm$ 2 & 2 $\pm$ 1 & 5.1 &  4.2 \\
        NGC 2960 & 140 $\pm$ 40 & 140 $\pm$ 40 & 0.9 $\pm$ 0.5 & 0.7 $\pm$ 0.3 & 3.4 &  3.1 \\
        NGC 3393 & 160 $\pm$ 50 & 150 $\pm$ 40 & 1.2 $\pm$ 0.6 & 0.6 $\pm$ 0.4 & 3.1 &  2.4 \\
        NGC 5495 & 120 $\pm$ 40 & 120 $\pm$ 40 & 7 $\pm$ 4 & 2 $\pm$ 2 & 4.4 &  3.3 \\
        NGC 5765b & 120 $\pm$ 50 & 120 $\pm$ 30 & 2 $\pm$ 1 & 2 $\pm$ 1 & 2.8 &  5.5 \\
\enddata
\tablecomments{Average values of polarization angle and fraction for our observed objects in both the $V$-band continuum (5800-6300\AA) and around H$\alpha$ (6500-6625\AA) \citep{ramos2016upholding} in the rest frame. The signal to noise ratio is quoted for the polarization fraction in the same ranges. The values of the polarization fraction should be considered lower limits because of galaxy continuum dilution in the intensity spectrum.}
\end{deluxetable*}

\subsection{Fitting the Spectra}
\label{sec:fit}

For each of the nine observed objects, we fitted the H$\alpha$-[NII] lines in both total and polarized intensity using the astropy LevMarLSQFitter function. In the total intensity, we used three Gaussian components for the narrow lines and a constant background continuum. The narrow lines were fixed to have the same width in velocity space. The relative amplitudes of the [NII] lines were fixed in a 1:3 ratio, and the relative positions were set by the known wavelength difference. In the polarized intensity, we fitted for the same three lines, although they are not always present in the polarized light, with the same constraints and a constant background. We also included a broad component represented by an additional Gaussian peak. The broad peak was initialized at the same wavelength as H$\alpha$, but its position was allowed to vary. In the case of NGC 1068, there are so many velocity components within our aperture that we could not find a model including narrow lines to fit the polarized spectrum well. Therefore, we only included the broad feature and constant background.

We are primarily interested in polarized intensity, and do not need a precise polarized fraction. Therefore, we do not perform starlight subtraction. We do note, however, that because we do not subtract the continuum, our polarization fractions should be considered lower limits.

\begin{figure*}[htb!]
	\plotone{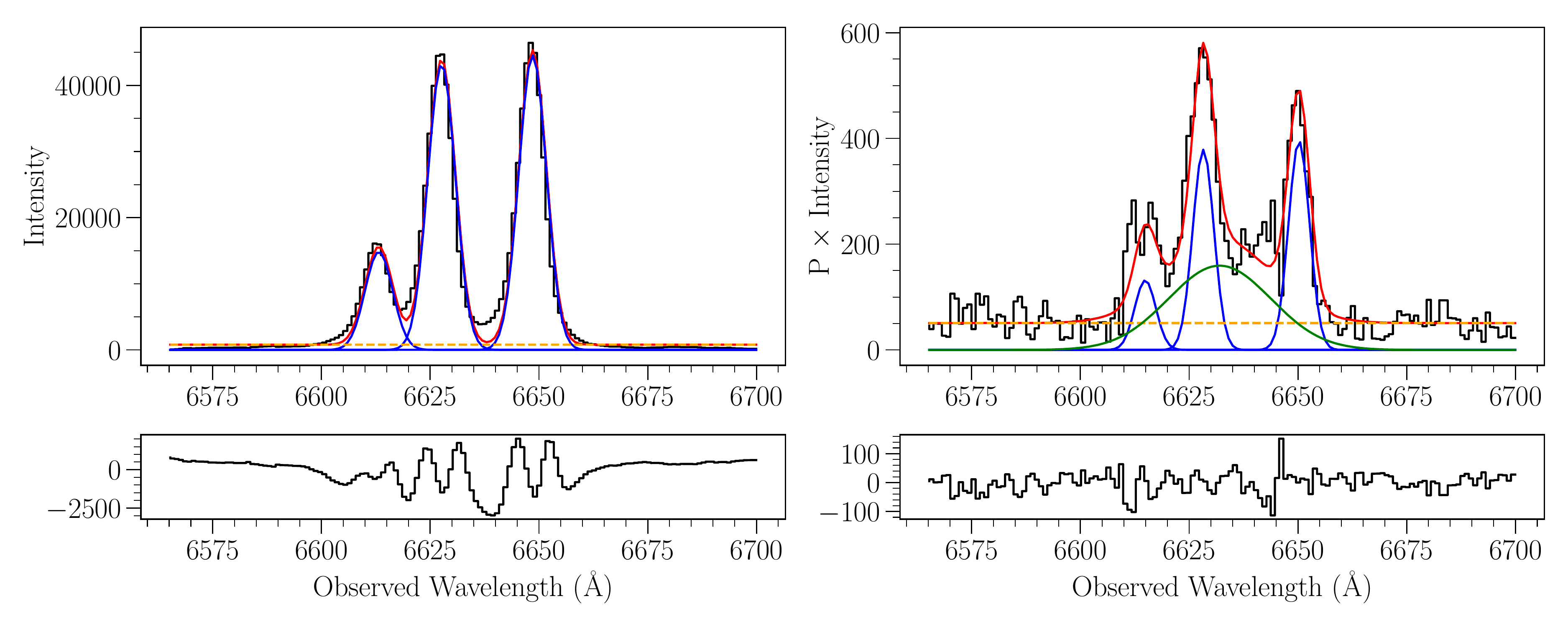}
    \caption{Spectra of IC 2560. The left panel shows the total intensity in arbitrary units along with the best fit. The narrow lines are represented in blue, the constant background with the yellow dashed line, and the total fit in red. The polarized intensity is given on the right. The fit includes an additional broad feature represented in green. The polarized spectrum is best fit with three narrow lines of width $\sim$ 260 km s$^{-1}$ and a broad component of width $\sim$ 1300 km s$^{-1}$.}
    \label{fig:ic2560_fit}
\end{figure*}

\begin{figure*}[htb!]
	\plotone{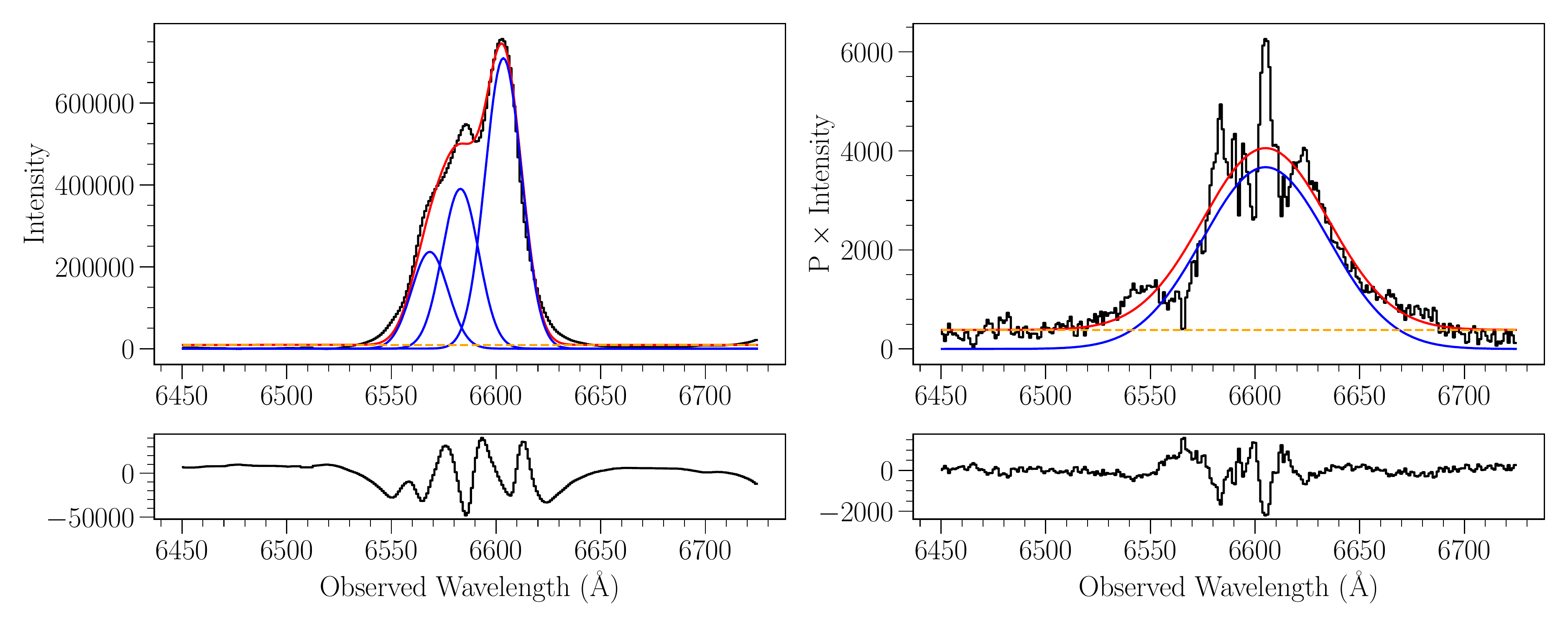}
    \caption{Same as Figure \ref{fig:ic2560_fit} but for NGC 1068. In polarized intensity, NGC 1068 only requires a broad component of width 3220 km s$^{-1}$ on top of a constant background.}
    \label{fig:ngc1068_fit}
\end{figure*}

\begin{figure*}[htb!]
	\plotone{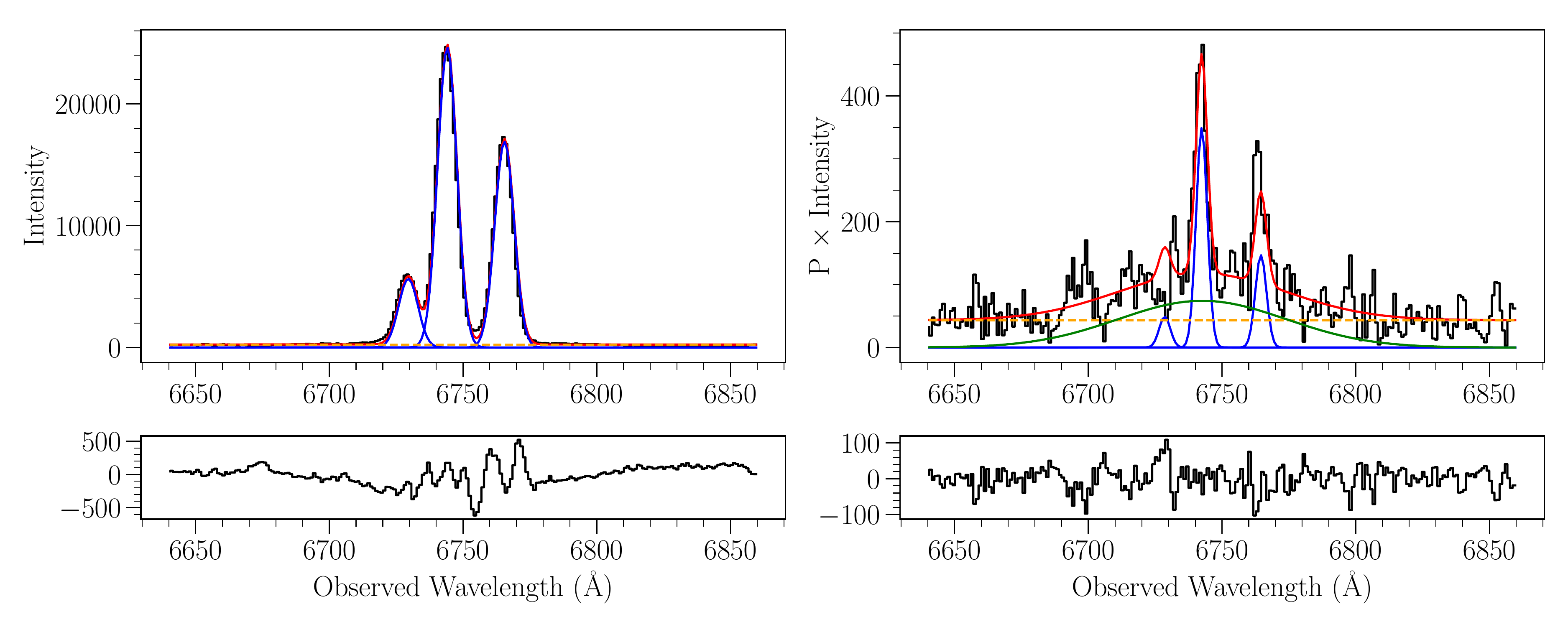}
    \caption{Same as Figure \ref{fig:ic2560_fit} but for NGC 5765b. The polarized intensity of NGC 5765b is best fit with narrow lines of width $\sim$ 220 km s$^{-1}$ and a broad component of width $\sim$ 3300 km s$^{-1}$.}
    \label{fig:ngc5765b_fit}
\end{figure*}

Of the nine objects, three show evidence of a broad H$\alpha$ feature: IC 2560, NGC 1068, and NGC 5765b. We find evidence of a broad feature in the weaker H$\beta$ line only for NGC 1068. The total and polarized fits for these objects along with residuals are shown in Figures \ref{fig:ic2560_fit} - \ref{fig:ngc5765b_fit}. 

To confirm the presence of a broad line in these three galaxies, we fitted the spectra with and without a broad component and calculated $\Delta \chi^2$ between the two models. For the broad component to be considered significant, $\Delta \chi^2$ must be greater than 2.7 (90\% confidence) times the number of additional parameters (three). We found $\Delta \chi^2$ to be greater than 8.1 in all cases.

To estimate the error in the broad line width, we took 1000 random samples from the spectrum assuming a normal distribution of polarized flux at each wavelength. The mean and standard error were assigned to be the observed values output by the reduction pipeline (\S 2.4). We then refit both the narrow and broad components in the artificial observations. Some samples did not require a broad line component by our $\Delta \chi^2$ test. We did not include these cases in calculating the distribution of broad-line properties. Any broad line with lower width than the narrow features was also excluded. Fewer than 15\% of samples were excluded by these conditions for each object. For each accepted sample, we determined the broad line parameters. From the distribution of broad-line widths, we found the difference from the mode to the 1$\sigma$ level of significance (the 84$^{\rm{th}}$ and 16$^{\rm{th}}$ percentiles, respectively). This provided the upper and lower errors on the FWHM measured from the original spectrum.

In addition, for each sample we calculated the percentage of total polarized flux contained by the broad line by integrating both the broad feature and the total polarized flux. The flux contained in the broad feature was not consistent with zero in any of the three objects, providing additional evidence that the broad feature is a significant component.

Observations of the remaining six objects were insufficient to confirm the presence of a broad feature. Although we do not find a polarized broad line in NGC 3393 with our observation, a broad line was found in this object by \cite{ramos2016upholding}.

The line width we find for the polarized broad line in NGC 1068 is narrower compared to previous results \citep[e.g.][]{antonucci1985spectropolarimetry, young1995near, inglis1994spatially}. We find a FWHM of 3220 $\pm$ 60 km s$^{-1}$ compared to 3750 $\pm$ 400 km s$^{-1}$ \citep{young1995near} and 4377 $\pm$ 300 km s$^{-1}$ \citep{inglis1994spatially}. This difference may be due to the variation of FWHM measured from different regions in the object, which may suggest a contribution from thermal broadening (see Section \ref{sec:linewidth}). For example, \cite{inglis1994spatially} measure a FWHM of 4377 $\pm$ 300 km s$^{-1}$ in the nucleus, but find a value of 3247 $\pm$ 400 km s$^{-1}$ at a location 2.5\arcsec\ NE. 

\begin{figure}[htb!]
	\plotone{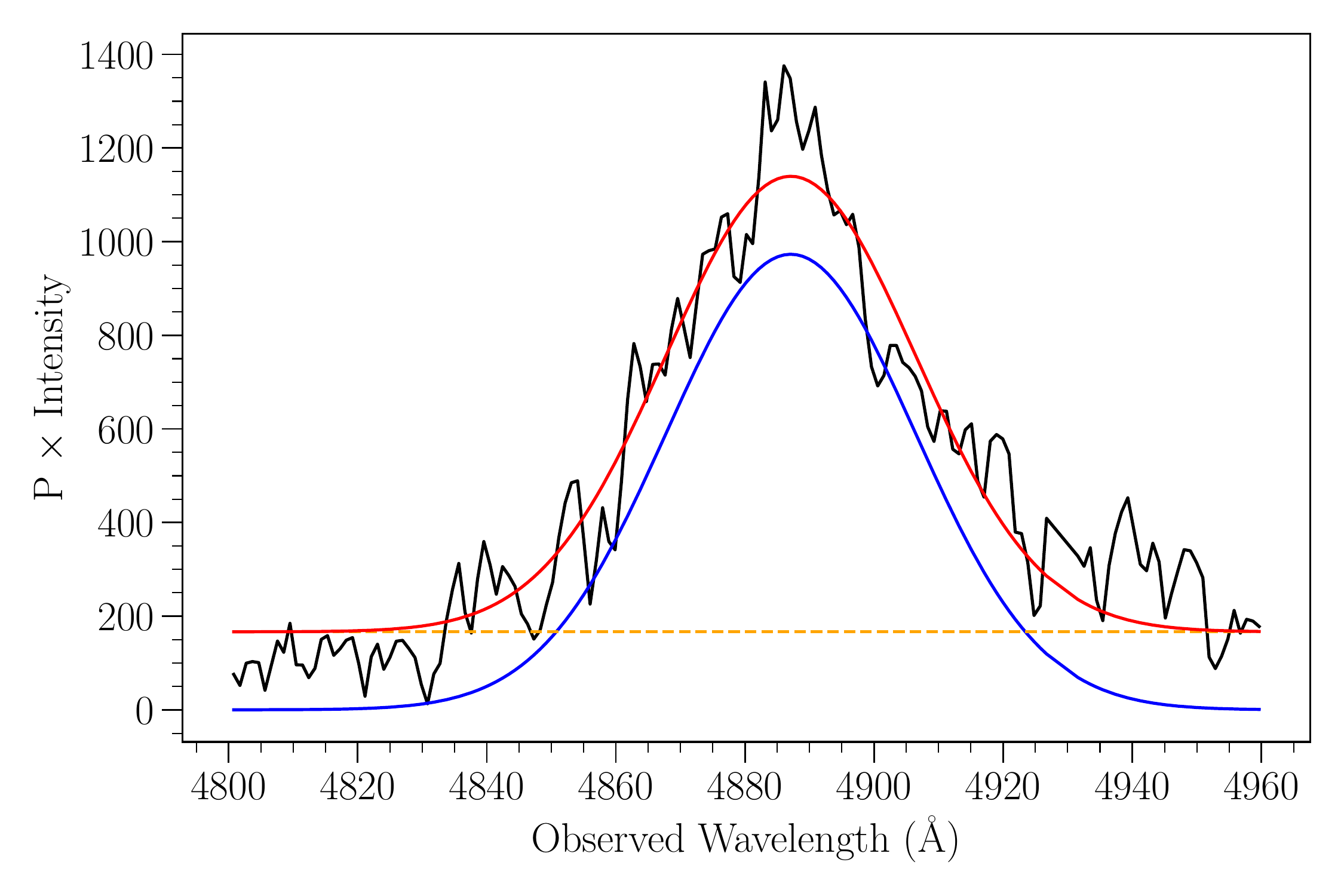}
    \caption{Fit of the polarized, broad H$\beta$ line in NGC 1068. The broad feature is fit with a Gaussian component (shown in blue) and a constant background (shown with the dashed orange line). We find the H$\beta$ line to have a width of $\sim2800\pm100$ km s$^{-1}$.}
    \label{fig:Hb}
\end{figure}

Additionally, while \cite{inglis1994spatially} find the FWHM of H$\beta$ to be 4290 $\pm$ 400 km s$^{-1}$ in the nucleus, \cite{miller1991multidirectional} measure 2900 $\pm$ 200 km s$^{-1}$ in a different region. We measure the H$\beta$ FWHM to be 2800 $\pm$ 100 km s$^{-1}$, as shown in Figure \ref{fig:Hb}. This is consistent with the  \cite{miller1991multidirectional} observation. Our H$\beta$ FWHM is narrower compared to our measurement of the broad H$\alpha$ line (3220 $\pm$ 60 km s$^{-1}$), possibly due to our lack of inclusion of the [NII] lines. We note that our value of FWHM has smaller error compared to the other observations. We are considering only the statistical error we estimated above rather than any systematic error from, for example, variations in FWHM between different regions.

\subsection{Broad Line Widths}
\label{sec:blwidths}

We combined our observed broad-line widths with additional objects from the literature. Here, broad-line width is defined using FWHM rather than line dispersion ($\sigma_{\rm{line}}$), which produces a different standard value of $f$ \citep[see e.g.][]{wang2019sloan}. We took all spectropolarimetric measurements with broad H$\alpha$ features in megamaser galaxies with known dynamical masses. The broad line widths are either provided directly from these sources, or in the case of NGC 2273 estimated by \cite{kuo2010megamaser} from the spectrum provided by \cite{moran2000frequency}.

We found additional values of FWHM for eight objects. For two of these objects, IC 2560 and NGC 1068, we also observed polarized broad features and took the mean of our measurements with those from the literature. Although we did not confirm the presence of polarized broad lines through our observations of NGC 3393, \cite{ramos2016upholding} do, so this object is included in our final sample. \cite{ramos2016upholding} perform a very similar fitting method to that described in Section \ref{sec:fit} on data from VLT/FORS2, and find broad features in four megamaser disk galaxies.

All measurements in our final sample are of polarized broad features, with the exception of NGC 4258. A broad feature in total intensity is observed by \cite{ho1997search} for this object. \cite{barth1999polarized} observe a polarized spectrum of NGC 4258, and fit the broad line fixing the width to the \cite{ho1997search} value. Although \cite{barth1999polarized} determine this is not a significant detection of a broad feature, we include this measurement because NGC 4258 is the archetypal megamaser galaxy and this is the only example of a broad line width for this object. Because \cite{barth1999polarized} fix the broad line width, we had to assign an uncertainty to the value. We chose a similar fractional error to the highest uncertainty measurements in our sample. 

In the case of NGC 4388, we have both a direct-light spectrum from \cite{ho1997search} and polarized spectrum \citep{ramos2016upholding}. \cite{ho1997search} observe a broad line with a FWHM of 3900 km s$^{-1}$ whereas \cite{ramos2016upholding} observe a polarized broad line with a FWHM of 4500$\pm$1400 km s$^{-1}$. Nominally these are consistent within the uncertainties for NGC 4388, but the difference is large, so we should view NGC 4258 with some skepticism.

All broad line widths including our observations and literature values are given in Table \ref{tab:fwhm}.

\section{Black Hole Masses}
\label{sec:BHmass}

Our main goal in this paper is to use the secure BH masses derived from the maser dynamics to test the fidelity of single-epoch BH masses. We thus review the strengths and limitations of each method briefly before turning to our comparison.

\subsection{Maser Dynamical Masses}

BH masses can be measured through the observation of dynamical tracers such as stars, gas, and masers \citep[e.g.][]{kormendy2013coevolution, mcconnell2013revisiting, saglia2016sinfoni}. Recently, CO emission has been used to measure BH mass \citep{davis2013black}, and with the use of ALMA these samples continue to grow \citep{boizelle2019precision}. Of these dynamical methods, the most precise extragalactic BH masses are determined with maser dynamics. There are significant limitations to this method, however. Megamaser disks must be edge-on to be detected and within $\sim100$ Mpc to be spatially resolved \citep{kuo2010megamaser}. To accurately determine the BH mass, the galaxies must also have a Keplerian rotation curve \citep{kuo2010megamaser}. 

\subsection{Reverberation Mapping}
\label{sec:RM}

While we cannot spatially resolve the broad line region, we can use temporal variability to determine a characteristic size. All AGN show variability in their disk emission on timescales of days to months, and because the BLR is photoionized by the UV photons from the accretion disk, this leads to variability in the broad line emission. There is a lag, however, because the BLR sits light-days from the BH. This time separation can be measured, and provides an estimate of the size scale of the BLR \citep[e.g.][]{blandford1982reverberation, peterson1993reverberation}. With the known speed of light, the average radius of the BLR can be determined. This can be used in combination with a velocity estimate and the virial parameter $f$ to calculate a virial mass (Equation \ref{eq:mass}).

\cite{onken2004supermassive} find $f$ to be a constant value of 1.4 when using FWHM as a velocity estimate. The value of $f$ is generally calibrated using the relationship between BH mass and galaxy stellar velocity dispersion ($\sigma_*$) by assuming that the scaling relations for quiescent galaxies are the same as those of AGN \citep[e.g.][]{onken2004supermassive, collin2006systematic, woo2010lick, graham2011expanded, grier2013stellar, batiste2017recalibration}. It is typically taken to be a constant of order unity, although it has been shown to vary between objects \citep{yu2019calibration} and there is an uncertainty of $\sim$0.4 dex on the value from calibration alone \citep{shen2013mass}. Any dependence of the scale factor $f$ on AGN properties (e.g. luminosity) constitutes a major systematic uncertainty in BH mass determination (\S \ref{sec:disc}).

High cadence, high signal-to-noise RM data allow for modeling of the BLR, as it becomes possible to measure the lag between continuum and line emission as a function of velocity. These data can constrain idealized models of the BLR that include flattening, inclination, and kinematic structure as free parameters, \citep[e.g.][]{pancoast2014modelling, grier2017structure, williams2018lick, williams2020space, bentz2021detailed, Villafa_a_2022}. In practice, simulated line profiles are generated from BLR models and then compared to the observed, reverberation mapping data to produce constraints on the model parameters. Using this technique, the mass of the BH is determined independently of the virial product, and therefore does not require a choice of $f$. Instead, the $f$-factor can be extracted as a model parameter along with the BH mass and other values of interest.

While the modeling is independent of $f$, it is limited by the models of the BLR that go into the fitting. If the models do not span the space of real BLRs in dynamics or structure, then it is possible to induce systematic errors in the BH masses. One interesting test is to attempt to recover BH masses from a simulated BLR. In an analysis of multiple RM modeling methods, \cite{mangham2019reverberation} use a rotating, biconical disk wind BLR model and generate mock data with simulations of ionization and radiative transfer. These data are passed through the RM modeling programs to evaluate how well they extract the BLR kinematics and parameters. Interestingly, while the CARAMEL model \citep{pancoast2011geometric, pancoast2014modellinga, pancoast2014modelling} fails to recover the proper kinematics, it does accurately recover the time delay, inclination of the BLR, and the BH mass.

We include RM modeling results for 16 Seyfert 1 sources from \cite{pancoast2014modelling}, \cite{grier2017structure}, and \cite{williams2018lick} as a comparison sample to our single-epoch maser measurements. We do not include additional RM modeled objects from \cite{williams2020space}, \cite{bentz2021detailed}, or \cite{Villafa_a_2022} as they do not calculate the virial parameter $f$.

\subsection{Single-Epoch Masses}
\label{sec:masscalc}

It is possible to use the radius-luminosity relation calibrated from RM to estimate a BH mass from a single spectrum, the so-called ``single-epoch" BH mass \citep{vestergaard2002determining}. Again, assuming that the BLR is virialized, one takes the luminosity and infers the size scale of the BLR using the radius-luminosity relation. As in RM, the velocity of the BLR gas comes from the line-width, and the virial factor $f$ is usually derived as a constant scaling that makes the ensemble of RM masses obey the M-$\sigma_*$ relation. Single-epoch masses typically have an uncertainty of $\sim$ 0.5 dex \citep{shen2013mass}. Here, our goal is to test these single-epoch masses against the well-known maser mass using our polarized broad line measurements

We estimate the virial products of our observed sample with Equation \ref{eq:vp}. We measure the velocity scale, represented by W in Equation \ref{eq:vp}, with the FWHM of the polarized broad lines, see Table \ref{tab:fwhm}. The size of the BLR is estimated through the radius-luminosity relation determined from RM, and is given by Equation \ref{eq:r_l} \citep{bentz2013low}.

\begin{eqnarray}
\label{eq:r_l}
\rm{log}_{10}(R_{BLR}/1 \textrm{ lt-day}) = 1.527^{+0.031}_{-0.031} + \nonumber \\ 0.533^{+0.035}_{-0.033} \rm{log}_{10}(\lambda L_\lambda / 10^{44} \textrm{ erg/s})
\end{eqnarray}

This relation gives the radius of the broad line region as a function of its 5100 \AA\ luminosity. Although recent results point to possible variation in the slope of the R-L relation \citep{alvarez2020sloan}, given the limited luminosity range of our sample, any slope variation will be minimal. The optical luminosity of the BLR cannot be measured directly in obscured AGN, so instead we choose to use high energy ($E >10 \, \rm{keV}$) X-rays to provide a proxy for the bolometric luminosity. Hard X-rays are highly penetrating even in the most Compton-thick AGN, which several of the masers sample are known to be \citep[e.g.][]{masini2016nustar}.

The most sensitive, and currently only, focusing hard X-ray telescope is the Nuclear Spectroscopic Telescope Array (NuSTAR), which provides high-quality X-ray spectroscopy in the 3--79 keV band. All nine of the masers considered here have previous observations with NuSTAR \citep{arevalo20142, balokovic2014nustar, bauer2015nustar, masini2016nustar, masini2019measuring}. These studies self-consistently fit each individual NuSTAR spectrum with a well-motivated transmission and reflection model to fully account for even the heaviest obscuration along the line of sight, providing the most direct measure of the intrinsic X-ray luminosity in the 2--10 keV~band derivable directly from the high-energy emission. We convert the intrinsic 2--10~keV luminosities provided by these studies to the luminosity distances adopted throughout (see Table \ref{tab:fwhm}). To estimate the bolometric luminosity for each maser, we use these hard X-ray derived 2--10 keV luminosities and adopt the luminosity-dependent bolometric correction of \cite{duras2020universal}, Equation \ref{eq:bol1}.

\begin{equation}
\label{eq:bol1}
K_X(L_X) = 15.33 \Bigg[1 + \bigg(\frac{\rm{log}(L_x/L_\odot)}{11.48}\bigg)^{16.20}\Bigg]
\end{equation}

Here, $L_X$ is the 2-10 keV intrinsic X-ray luminosity. The error on $K_X$ is dominated by the intrinsic scatter of 0.37 dex. \cite{duras2020universal} provides an additional bolometric correction for 4400 \AA\ luminosity, Equation \ref{eq:bol2}. 

\begin{equation}
\label{eq:bol2}
K_O(L_{\rm{BOL}}) = 5.13
\end{equation}

Again, the error is dominated by intrinsic scatter with a value of 0.26 dex. To convert from 4400 \AA\ to 5100 \AA\ luminosity, we use the power law fit to the composite spectrum in \cite{berk2001composite}, Equation \ref{eq:lum}.

\begin{equation}
\label{eq:lum}
f_\lambda \propto \lambda^{-1.56}
\end{equation}

This gives the 5100 \AA\ luminosity to be approximately 80\% of that at 4400 \AA. The final luminosity values can then be used in Equation \ref{eq:r_l} to calculate the radius of the broad line region.

With these values, the virial products of each object can be estimated, and are given in Table \ref{tab:fwhm}. The virial products can then be compared to dynamical masses from maser disks  \citep{greene2016megamaser, kuo2020megamaser}.

\begin{deluxetable*}{lcccccccccc}
\tablecaption{\label{tab:fwhm} Megamaser Sample}
\setlength\tabcolsep{1.6pt}
\renewcommand{\arraystretch}{1.1}
\tablehead{
\colhead{Galaxy} &  \colhead{H$\alpha$ FWHM} & \colhead{Ref.} & \colhead{Avg H$\alpha$ FWHM} & \colhead{D} & \colhead{log M$_{\rm{BH}}$} & \colhead{Ref.} & \colhead{log L$_{2-10}$} & \colhead{$\lambda_{edd}$} & \colhead{log Virial Product} & \colhead{log(f)}\\ & \colhead{(km s$^{-1}$)} &  & \colhead{(km s$^{-1}$)} & \colhead{(Mpc)} &  \colhead{(M$_\odot$)} & & \colhead{(erg/s)} & & \colhead{(M$_\odot$)}
}
\decimalcolnumbers
\startdata
Circinus  & 2300 $\pm$ 500 & 1 & 2300 $\pm$ 500 & 2.8 & 6.06 $\pm$ 0.1 & 9 & 42.2 & -0.8 & 6.8 $\pm$ 0.3& -0.7 $\pm$ 0.3\\
IC 2560  & 1300 $\pm$ 100 & 2 & 1700 $\pm$ 200 & 41.8 & 6.64 $\pm$ 0.06 & 9 & 43.4 & -0.1 & 7.2 $\pm$ 0.3& -0.5 $\pm$ 0.3\\
& 2100 $\pm$ 300 & 1\\
Mrk 1210  & 2380 $\pm$ 120 & 3 & 2380 $\pm$ 120 & 56.7 & 7.152 $\pm$ 0.006 & 10 & 43.3 & -0.8 & 7.4 $\pm$ 0.3& -0.2 $\pm$ 0.3\\
NGC 1068  & 3220 $\pm$ 60 & 2 & 3800 $\pm$ 200 & 15.9 & 6.92 $\pm$ 0.25 & 9 & 43.4 & -0.4 & 7.9 $\pm$ 0.3& -1.0 $\pm$ 0.4\\
& 3750 $\pm$ 400 & 4\\
& 4377 $\pm$ 300 & 5\\
NGC 2273  & 2900 $\pm$ 200 & 6,7 & 2900 $\pm$ 200 & 25.7 & 6.88 $\pm$ 0.02 & 10 & 43.0 & -0.8 & 7.4 $\pm$ 0.3& -0.5 $\pm$ 0.3\\
NGC 3393  & 5000 $\pm$ 600 & 1 & 5000 $\pm$ 600 & 49.2 & 7.2 $\pm$ 0.33 & 9 & 43.3 & -0.8 & 8.0 $\pm$ 0.3& -0.8 $\pm$ 0.4\\
NGC 4258  & 1700 $\pm$ 500 & 8 & 1700 $\pm$ 500 & 7.3 & 7.58 $\pm$ 0.03 & 9 & 41.2 & -3.3 & 6.0 $\pm$ 0.4& 1.6 $\pm$ 0.4\\
NGC 4388  & 4500 $\pm$ 1400 & 1 & 4500 $\pm$ 1400 & 19.0 & 6.92 $\pm$ 0.01 & 10 & 42.6 & -1.2 & 7.6 $\pm$ 0.4& -0.7 $\pm$ 0.4\\
NGC 5765b  & 3300$_{-300}^{+500}$ & 2 & 3300$_{-300}^{+500}$ & 126.3 & 7.66 $\pm$ 0.04 & 10 & 43.0 & -1.5 & 7.6 $\pm$ 0.3& 0.1 $\pm$ 0.3\\
\enddata
\tablecomments{Column 1: Object name. Columns 2-3: FWHM of the H$\alpha$ broad line and reference. Column 4: Averaged FWHM of H$\alpha$ broad line. Columns 5-7: Distance and dynamical black hole mass with references. Values are taken from the sources listed in each reference. Column 8: Instrinsic 2-10 keV X-ray luminosity. All values have estimated error of 0.1 dex except for NGC 1068 and NGC 1194 which have uncertainties of 0.3 dex. Column 9: Eddington ratio estimated from bolometric luminosity (see Equation \ref{eq:bol1}) and dynamical BH mass. Column 10: Virial product calculated with Equation \ref{eq:vp}. Column 11: $f$-value estimated by comparing dynamical mass and virial product. References: (1) \cite{ramos2016upholding}, (2) Our Work, (3) \cite{tran1995naturea,tran1995natureb}, (4) \cite{young1995near}, (5) \cite{inglis1994spatially}, (6) \cite{moran2000frequency} (spectrum), (7) \cite{kuo2010megamaser} (value), (8) \cite{ho1997search}, (9) \cite{greene2016megamaser}, (10) \cite{kuo2020megamaser}}
\end{deluxetable*}

\begin{figure*}[htb!]
	\plotone{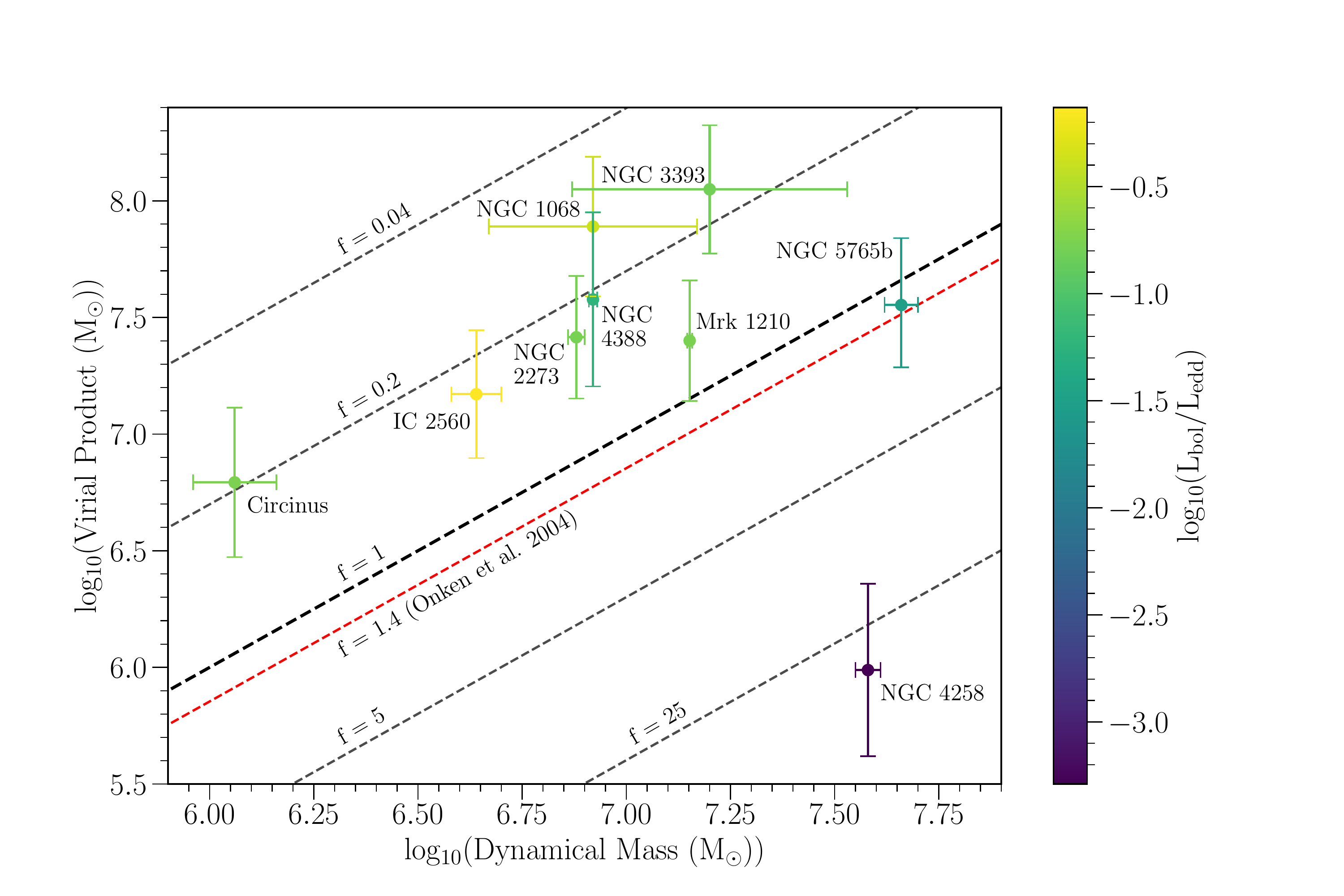}
    \caption{Comparison between known dynamical mass and the virial product without including an $f$-factor. Each object is colored by its Eddington ratio. The diagonal dashed lines represent the predicted BH masses from virial product using different values of $f$. The center, bold line has $f$ = 1, and each line below (above) increases (decreases) $f$ by a factor of 5. The red dashed line represents a standard value of $f$ = 1.4 calibrated with FWHM \citep{onken2004supermassive}.}
    \label{fig:masscomp}
\end{figure*}

In Figure \ref{fig:masscomp}, we compare the virial product of each maser to its known dynamical mass. These products do not include the $f$-factor. Rather, different values of $f$ are represented by the dashed lines in the figure. If all objects were to fall on the center, bold line, the BH mass would be equal to the virial product with no additional geometric factor. Each additional line has an $f$ value that is five times higher than the line above it. Therefore, within this maser sample $f$ ranges from approximately 0.1 to 40 (or 0.1 to 1.3 if NGC 4258 is excluded).

For a measured value of the virial product, the true dynamical mass of the object can vary greatly. Sources with a virial product of approximately $10^7$ M$_\odot$, for example, can have a dynamical mass from $10^6$ - $10^8$ M$_\odot$ within only this sample of nine objects. Therefore, $f$ must be well calibrated to determine the true dynamical mass from the virial product. We will consider relationships between $f$ and other observable parameters in Section \ref{sec:corr}. We will also explore whether we should expect to find a well-defined $f$ factor given the uncertainties in our virial product.

\section{Discussion}
\label{sec:disc}

We first discuss theoretical models of the BLR, and the importance of different measured parameters. Then we turn to our final sample, which includes 9 masers with broad polarized lines and 16 RM+dynamical modeling objects from the literature. We consider whether the single-epoch masses are correlated with the dynamical mass measurement, first with the maser sample alone, where we understand well the dynamical masses and uncertainties, then including the full sample. We conclude with a discussion of the caveats to these results.

\subsection{Theoretical expectations for BLR structure}

We have focused on using the BLR as a tracer of the BH mass. The BLR, however, is also one of our primary tools for understanding accretion onto supermassive BHs, as the ionization structure gives us clues about the SED of the accretion disk, and the dynamics are tied to the emission from the disk. Over many decades, several different models have been proposed for the BLR, including orbiting clouds, inflowing and outflowing gas, and rotating disk winds \citep[see reviews in][]{mathews1985structure, sulentic2000phenomenology, Czerny2019}, each of which show some success, particularly in explaining the photoionization of the BLR gas \citep[e.g.][]{kwan1981formation, korista2004optical}. RM is one of the best ways to probe this region, and velocity resolved RM is now within reach for large samples.

One of the models that has been explored most is that of the disk wind \citep[e.g.][]{shields1977origin, emmering1992magnetic, chiang1996reverberation} where the gas is accelerated by line driving. Although x-ray emission may reduce the efficiency of line-driven winds \citep{waters2016reverberation}, shielding near the central BH ultimately enables line-driving to occur \citep{proga2000dynamics}. These models naturally explain observations like high velocity absorption lines observed in quasars, the observed BLR line ratios \citep{chiang1996reverberation}, and even the echo images resulting from RM campaigns \citep{waters2016reverberation}. The disk wind models are also consistent with measurements of line kinematics and line strengths from the BLR \citep[e.g.][]{proga2004dynamics}. 

In the disk wind model, there are several concrete predictions for how $f$ might depend secondarily on other parameters. For instance, we expect the geometry of the system, including inclination and relative positions of the BLR and polar scatterers, to affect observed line width, an effect which must be accounted for in $f$ \citep[e.g.][]{chiang1996reverberation, smith2002spectropolarimetric, proga2004dynamics, waters2016reverberation}. \cite{proga2004dynamics} find the disk wind to be sensitive to Eddington ratio. In general, the disk structure and temperature depend on luminosity, and therefore too the BLR, so there is a strong motivation to explore this question empirically.

\subsection{Correlations between observable and derived values}
\label{sec:corr}

For the maser sample as a whole, we do not find clear evidence of a correlation between virial product and dynamical BH mass. We will quantify the lack of correlation below. However, in this section we also try to determine whether there is a secondary parameter driving the relation between $f$ and dynamical BH mass, which might help us understand the virial BH masses.

We evaluate the possible trend between the virial product and dynamical mass, along with other correlations, using the Pearson correlation test. Because the values we will compare have individual, possibly correlated, errors we use a Monte Carlo (MC) method to explore the correlations rather than taking the Pearson correlation as measured. To perform the MC correlation test, 10,000 samples of BLR radius (or luminosity), FWHM, and dynamical mass are taken assuming a normal distribution for each with the mean and standard error set by values in Table \ref{tab:fwhm} and Equations \ref{eq:r_l}-\ref{eq:lum}. These values are then used to recalculate the virial product, and we calculate $f$ by dividing the dynamical mass by the virial product. We measure the $r$ and $p$ values from the Pearson test on each sample and look at the distributions of $r$ and $p$ over all samples to evaluate the correlation between different values.

The distributions of these values for the relationship between virial product and the mass of the BH are given in Table \ref{tab:corr}. Additional correlations are given in Table \ref{tab:app_corr} in Appendix B. If the virial product is a good estimator for the dynamical mass of the BH, there should be a high correlation between the two values. In the full sample of masers, however, we see no correlation between the mass of the BH and the virial product. We additionally see no correlation between the mass of the BH and the individual components of the virial product: luminosity or BLR radius and FWHM. Even if NGC 4258 is removed, there is no significant correlation. This implies that the $f$-factor is unlikely to be a constant value, and that the BH mass depends sensitively on the per-object value of $f$. Given that $f$ is not independent of dynamical mass, we do recover an expected trend between implied per-object $f$-value and M$_{\textrm{dyn}}$ for the maser sample.

\begin{deluxetable*}{llcc}
\tablecaption{\label{tab:corr} Correlation Test Results}
\tablewidth{0pt}
\renewcommand{\arraystretch}{1.3}
\tablehead{
\colhead{Objects} & \colhead{Comparison} & \colhead{$r$} & \colhead{$p$}
}
\decimalcolnumbers
\startdata
        All Masers & $\rm{M_{BH}}$ - Virial Product & $-0.05_{-0.20, -0.34}^{+0.24, +0.49}$ & $0.68_{-0.27, -0.48}^{+0.22, +0.30}$\\
        \hline
        No NGC 4258 & $\rm{M_{BH}}$ - Virial Product & $0.15_{-0.24, -0.42}^{+0.29, +0.58}$ & $0.64_{-0.37, -0.60}^{+0.25, +0.34}$\\
        \hline 
        Masers and RM Modeling & $\rm{M_{BH}}$ - log$_{10}$f & $0.46_{-0.11, -0.23}^{+0.10, +0.18}$ & $0.02_{-0.02, -0.02}^{+0.06, +0.25}$\\
        & log$_{10}$f - L & $0.07_{-0.08, -0.15}^{+0.08, +0.15}$ & $0.72_{-0.24, -0.44}^{+0.20, +0.27}$\\
        & log$_{10}$f - FWHM & $-0.36_{-0.08, -0.14}^{+0.08, +0.16}$ & $0.08_{-0.05, -0.07}^{+0.10, +0.27}$\\
\enddata
\tablecomments{Selected results of Pearson's $r$ test. The $r$ and $p$ values shown are the median along with bounds containing 68 and 95\% of the random samples. The full correlation test results are given in Appendix B.}
\end{deluxetable*}

If a relationship existed between the observed parameters, i.e. luminosity and FWHM, with the $f$-factor, these values could be used to calibrate $f$ and find an accurate mass from the virial product. Additionally, from our understanding of the structure of the BLR, we may expect a dependence on luminosity. Therefore, we search for a correlation between $f$ and luminosity or FWHM, but do not find a strong correlation with either. We similarly would expect a relationship between Eddington ratio and $f$-factor, and we observe a possible correlation in Figure \ref{fig:masscomp}. When the Pearson test is performed, however, we do not see a strong relationship between these parameters.

\begin{figure*}[htb!]
	\plotone{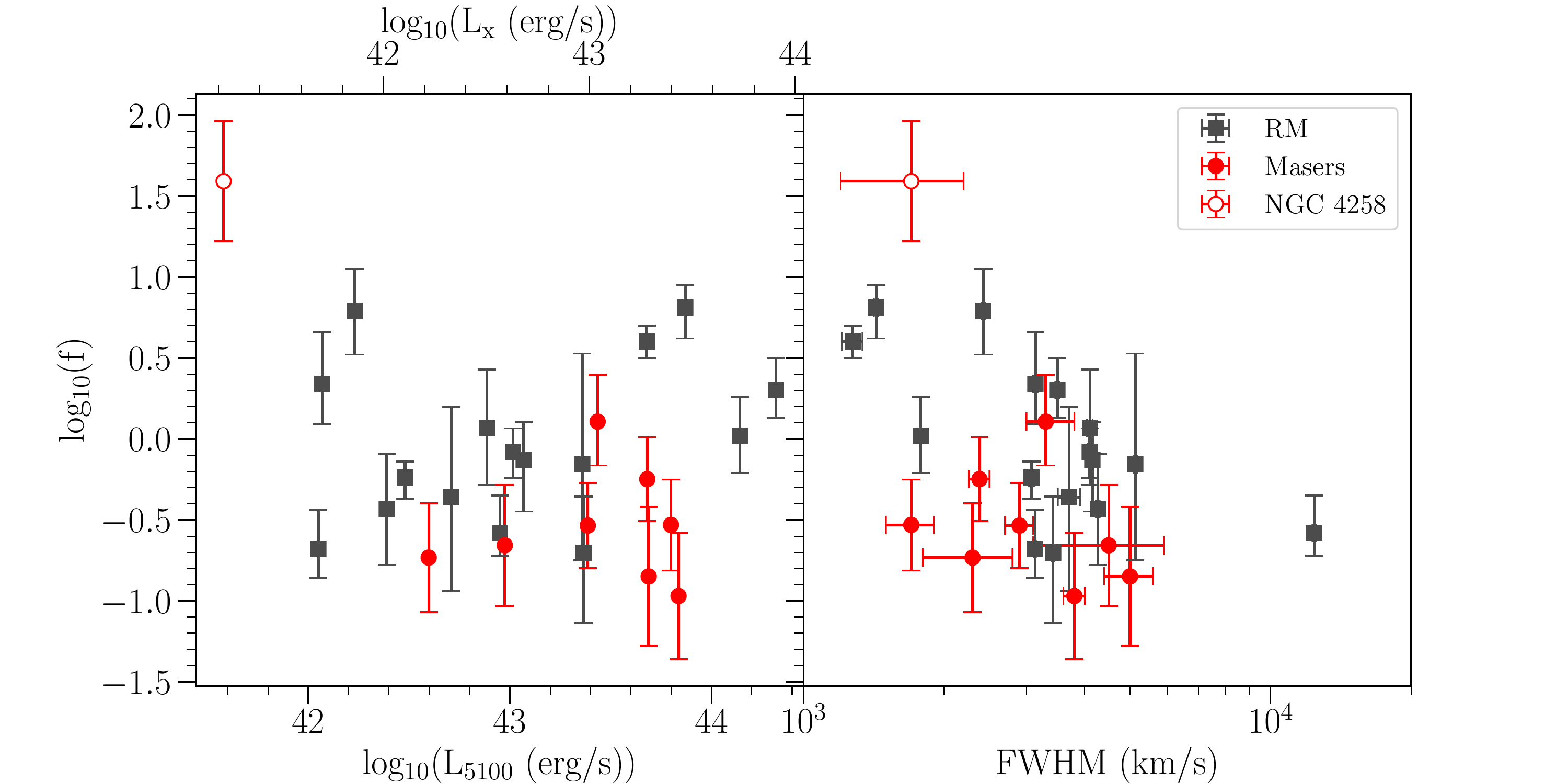}
    \caption{Values of the $f$-factor for both the megamaser and modeling samples as compared to BLR luminosity and width of the polarized broad line. NGC 4258 is represented by an open circle due to the issues with its measurement. Values of maser luminosity have uncertainties of $\sim$0.5 dex. FWHM values for the objects in \cite{pancoast2014modelling} and \cite{williams2018lick} are taken from \cite{park2012lick} and \cite{barth2015lick} respectively and correspond to the H$\beta$ line.}
    \label{fig:threepanel}
\end{figure*}

\begin{figure*}[htb!]
	\plotone{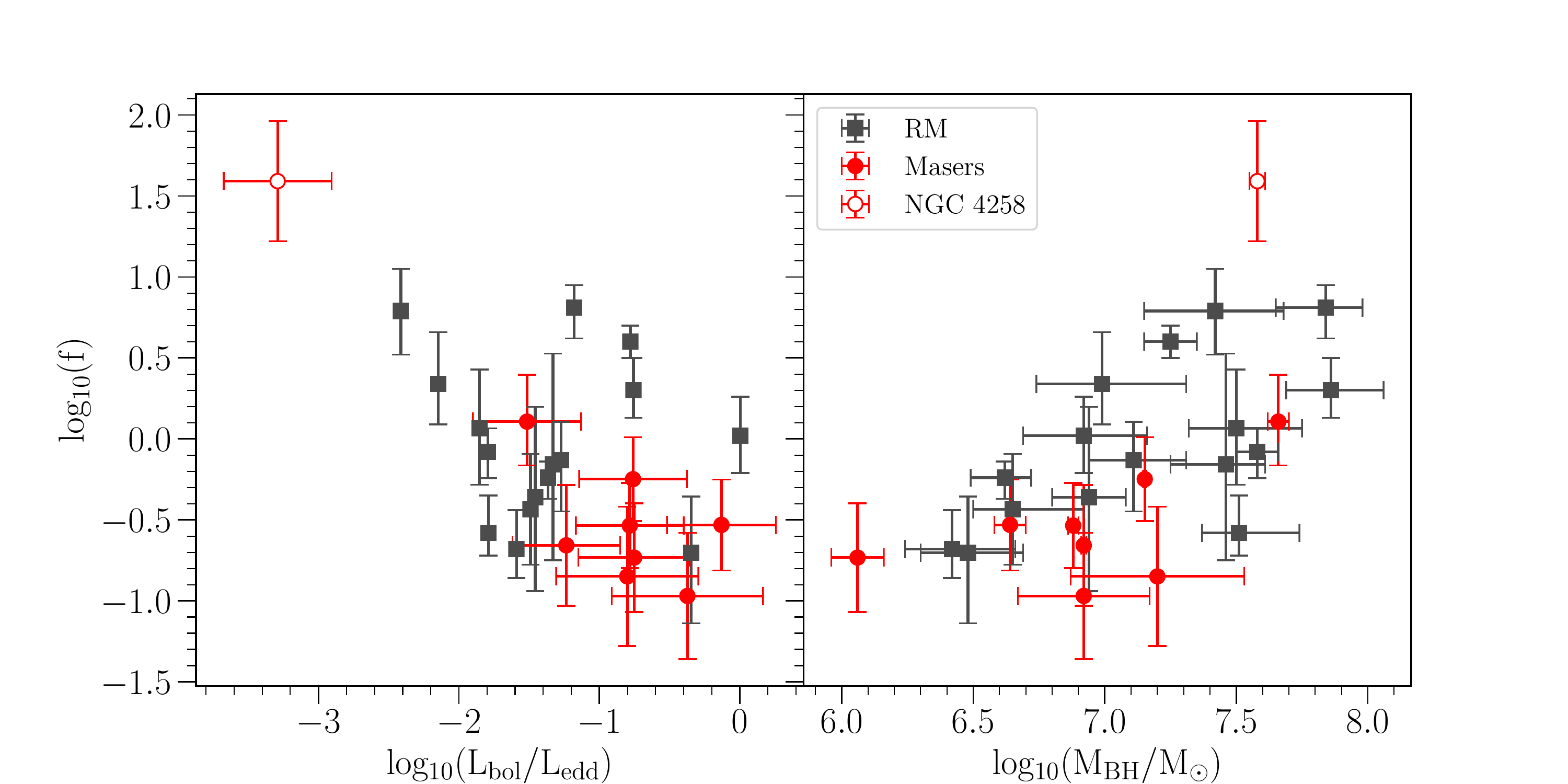}
    \caption{Values of the $f$-factor for both the megamaser and modeling samples as compared to the Eddington ratio and dynamical mass of the objects.}
    \label{fig:twopanel}
\end{figure*}

After exploring the correlations between parameters using only the masers, we combine the maser sample with the RM sample described in Section \ref{sec:RM}. Figures \ref{fig:threepanel} and \ref{fig:twopanel} show comparisons between $f$ and observable and derived values, respectively, including all objects. The two panels of Figure \ref{fig:threepanel} show $f$ compared to luminosity and FWHM respectively. We find no evidence of a correlation between these values.

The $f$-factor does appear to have a relationship with the Eddington ratio and BH mass when including the maser and RM samples, as seen in Figure \ref{fig:twopanel}. Considering only the maser sample, we would expect $f$ to be related to Eddington ratio and dynamical mass because the mass of the BH is included in all three values; $f$ is directly proportional to the dynamical mass and Eddington ratio is inversely related. We note that we still see this possible relationship in the combined sample, where $f$ is not derived directly from M$_{\rm{BH}}$ as in the masers. The Pearson test, however, does not provide evidence for a strong relationship between these parameters. Additionally, even if there was a relationship between $f$, Eddington ratio, and dynamical mass, these values are not directly measurable and therefore would not be useful for determining the value of $f$ for a given object. 

In the past, one confirmation that single-epoch measurements are accurate tracers of BH mass has been the measured correlation between $\sigma_*$ and virial product. Indeed, $f$ has been calibrated by solving for the value bringing the virial products in line with the M-$\sigma_*$ relation. Therefore, we also examine the relationship between BH mass, $\sigma_*$, and $f$. Because the M-$\sigma_*$ relationship is used to calibrate $f$, we should see a correlation between $\sigma_*$ and BH mass or $f$. We compile values of $\sigma_*$ for the majority of the objects in our sample, see Table \ref{tab:app_big} in Appendix C. When performing the correlation test, however, we do not find evidence for a relationship between $\sigma_*$ and BH mass or $f$ in either the sample of megamasers taken alone or when RM modeling objects are included.

Given the large uncertainties on individual virial products, it is important to ask whether we could measure a single $f$ value from our sample even if virial product and dynamical mass were perfectly correlated. We must test if our sample is strong enough to rule out a relationship between virial product and dynamical mass, and do so as follows.

\begin{figure*}[htb!]
	\plotone{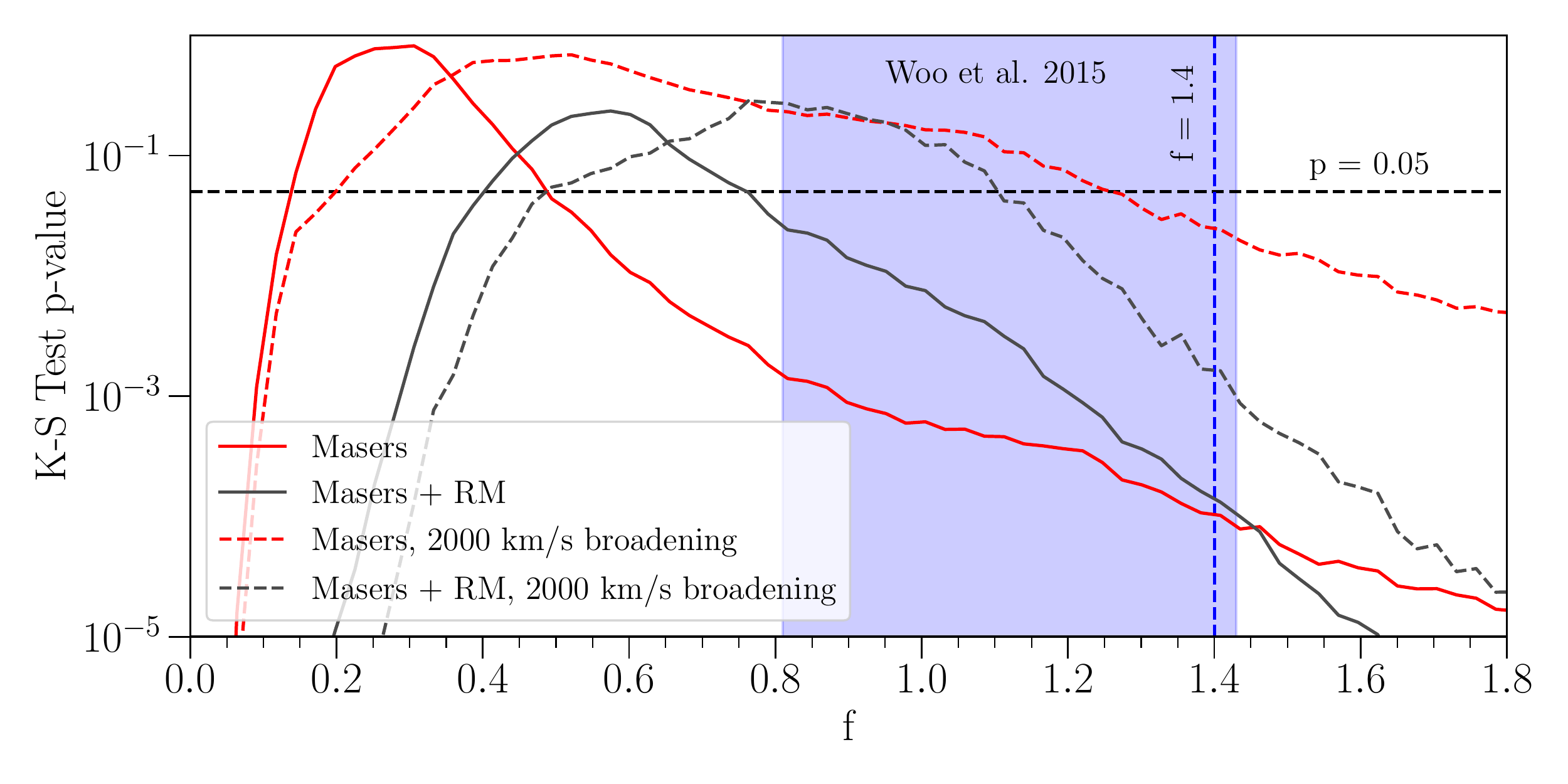}
    \caption{K-S test p-value generated by comparing the observed maser or maser and RM modeling sample to a randomly generated, perfectly correlated sample. The blue shaded region represents the values of $f$ included in the 1$\sigma$ range of \cite{woo2015black}, $f = 1.12 \pm 0.3$. The vertical blue line gives $f$ = 1.4 as predicted by \citep{onken2004supermassive}. The horizontal line represents a $p$ value of 0.05. If a value of $f$ falls below the line, we can reject a perfect correlation between dynamical mass and virial product with that single $f$ value. The test is performed for virial products calculated with the values of FWHM given in Table \ref{tab:fwhm} (solid lines), and the values of FWHM after 2000 km s$^{-1}$ broadening is subtracted in quadrature (dashed lines). We find it to be unlikely that our unchanged sample could be drawn from a perfect correlation between dynamical mass and virial product with $f$ from either \cite{woo2015black} or \cite{onken2004supermassive}. It would be more difficult, however, to reject these $f$ values if thermal broadening were present.}
    \label{fig:ks}
\end{figure*}

First, we generate artificial sources by selecting dynamical masses from a uniform distribution spanning the range of our maser sample. Assuming a perfect correlation between dynamical mass and virial product, we calculate virial products by choosing a constant value of $f$. We assign errors to both quantities by taking the average error associated with the virial products and dynamical masses in our maser sample. This allows us to simulate an observation of each object by sampling a Gaussian with mean given by the generated virial product or dynamical mass, and a standard error from the average value. We repeat the process to generate a random sample with $10^4$ simulated maser objects, each with an observed dynamical mass and virial product. These values are combined to generate an observed $f$ value. The Kolmogorov–Smirnov (K-S) test is used to compare the generated sample to our true sample of nine maser objects. This is done for different values for $f$, and including both the masers alone and the combined sample of masers and RM modeling objects. Results are shown as the solid lines in Figure \ref{fig:ks}.

From the results of the K-S test, we can rule out a correlation between dynamical mass and virial product with a single value of $f$ = 1.4 \citep{onken2004supermassive}. We can also almost entirely reject the 1$\sigma$ range of $f = 1.12 \pm 0.3$ given by \cite{woo2015black} with a $p$-value below 0.05. However, we cannot rule out a perfect correlation for all values of $f$. For example, when including the maser sample alone, values of $f$ between $\sim$0.1-0.5 are allowed. As seen in Figure \ref{fig:masscomp}, the majority of maser objects fall in this range of $f$ values. To show that virial product and dynamical mass are not correlated with any value of $f$, we would need to observe additional maser objects or reduce the uncertainties on existing objects. 

The major sources of error in our virial product estimates are our understanding of the polarized broad line width and measurements of the intrinsic luminosity. These will be discussed in the following sections. Reducing these uncertainties will be difficult, so we can instead estimate how many additional maser objects would be required to rule out a single value of $f$. Starting with the observed sample of masers, we assume that there is no relationship between virial product and dynamical mass. We generate a random value of each from a uniform distribution over our sample range. Using these two values, we calculate $f$. We perform the K-S test again with this additional object included in the maser sample, and continue adding objects until $p < 0.05$. We repeat the test 1000 times and fit the resulting distribution with a Poisson function to statistically determine how many additional objects are required. For a fixed $f$ of 0.3, we find that approximately six additional maser objects would be required to rule out a correlation.

\subsection{Caveats from intrinsic luminosity}

The requirement of converting a hard X-ray luminosity in a heavily obscured AGN to optical continuum measurement introduces several sources of error. As discussed in Section \ref{sec:masscalc}, there is significant uncertainty in the bolometric corrections. The X-ray correction has an intrinsic scatter of 0.37 dex, while the optical has a scatter of 0.26 dex \citep{duras2020universal}. Future surveys that attempt to measure virial products across all types of AGN could potentially reduce the combined uncertainty on the luminosity through the consistent calibration to a wavelength region that is relatively insensitive to dust and gas obscuration, such as using high spatial resolution imaging in the mid-infrared.

\subsection{Caveats in using polarized line widths}
\label{sec:linewidth}

We have presented comparisons between dynamical mass and virial products using broad-line widths measured from polarized light. We must address whether the polarized line widths may either under or overestimate the true velocity distribution. 

In megamaser galaxies, the disk must be nearly edge-on to observe masing. When jets are seen, they align with the angular momentum of the disk, suggesting the masers and inner disk are aligned \citep{kamali2019accretion}. With this geometry we would expect polar scattering, where scatterers are located above and below the disk, to dominate. If this were the case, the polarization angle is predicted to be perpendicular to the radio jet \citep[e.g.][]{smith2002spectropolarimetric}. For the objects in our sample which are observed to have a jet, we do find these two angles to be approximately perpendicular. We can make this comparison for eight objects from the maser sample, see Table \ref{tab:angle}. Polar scattering would produce narrower polarized lines compared to the full distribution of velocities in the BLR \citep[e.g.][]{smith2002spectropolarimetric, smith2004seyferts}. Therefore, the geometry of the object may lead to narrowing of the polarized broad lines. We expect the order of this effect to be comparable to the inclination effects for the BLR sample.

\begin{deluxetable}{lcccc}
\tablecaption{\label{tab:angle} Polarization and Jet Angles}
\tablewidth{0pt}
\tablehead{
\colhead{Object} & \colhead{$\theta$ (deg)} & \colhead{Ref.} & \colhead{Jet Angle (deg)} & \colhead{Ref.}
}
\decimalcolnumbers
\startdata
IC 2560 & 91 $\pm$ 1 & 1 & 24 & 6\\
Mrk 1210 & 29 & 2 & 125 & 7\\
NGC 1194 & 150 $\pm$ 30 & 3 & 56 & 8\\
NGC 2273 & 25 $\pm$ 7 & 4 & 90 & 9\\
NGC 2960 & 140 $\pm$ 40 & 3 & 125 $\pm$ 10 & 10\\
NGC 3393 & 2 $\pm$ 4 & 1 & 56 & 8\\
NGC 4258 & 79 $\pm$ 1 & 5 & -3 $\pm$ 1 & 11\\
NGC 4388 & 95 $\pm$ 9 & 1 & 24 & 12\\
\enddata
\tablecomments{Polarization angles measured in the continuum and radio jet position angles for objects in our maser sample with an observed radio jet. Angles are measured East of North. We find the these angles to be roughly perpendicular, with an unweighted, average separation of $\sim70\pm 20^{\circ}$. References: (1) \cite{ramos2016upholding}, (2) \cite{tran1995natureb}, (3) Our Work, (4) \cite{moran2000frequency}, (5) \cite{barth1999polarized}, (6) \cite{yamauchi2012water}, (7) \cite{xanthopoulos2010linear}, (8) \cite{schmitt2001jet}, (9) \cite{ulvestad1984radio}, (10) \cite{sun2013refining}, (11) \cite{cecil2000active}, (12) \cite{falcke1998hubble}.}
\end{deluxetable}

In addition to line narrowing due to the scattering geometry in these objects, we also expect some thermal broadening due to the nature of the scatterers themselves. Studies of NGC 1068 have shown evidence for scattering by both dust and electrons with the dominant scatterer varying by region \citep[e.g.][]{miller1991multidirectional}. By comparing the observed polarized line width between dust scattering regions, which are not expected to cause thermal broadening, and electron scattering areas, which do produce broadening, the effect of thermal broadening can be estimated. NGC 1068 was found to have thermal broadening of approximately 3360 km s$^{-1}$ with a corresponding electron temperature of $\sim10^5$ K \citep{miller1991multidirectional}.

The thermal broadening in NGC 1068 likely represents an extreme case. First, the broadening of $\sim3400$ km s$^{-1}$ is larger than the total line width of the majority of objects in our sample. Additionally, regions dominated by dust or electron scattering could be observed independently in NGC 1068, while we mostly likely see a mix of both scatterers when observing the objects in our sample. We can consider NGC 4388 for which we have both direct-light and polarized broad line measurements with FWHMs of 3900 km s$^{-1}$ \citep{ho1997search} and 4500$\pm$1400 km s$^{-1}$ \citep{ramos2016upholding} respectively. If we take the difference between these values to be solely due to thermal broadening, we find a broadening of $\sim 2200$ km s$^{-1}$ with a corresponding electron temperature of $3 * 10^4$ K. This is significantly lower than the broadening in NGC 1068.

Although we do not expect all of our objects to have as much thermal broadening as NGC 1068, we may consider what effect more moderate broadening would have on our results. In general, our virial masses overestimate the dynamical masses of the BHs if a typical value of $f$ = 1.4 is assumed. It is possible that thermal broadening is responsible for an increase in the observed FWHM leading to this overestimation. Therefore, we determine the value of thermal broadening that must be subtracted in quadrature from the FWHM of each object such that the virial and dynamical masses agree for $f$ = 1.4. For this test, we exclude NGC 4258 which would require the FWHM to be narrower to agree. We find our sample to require an average thermal broadening of $\Delta v \approx 3000 \pm 1000$ km s$^{-1}$ to match the dynamical masses using the accepted value of the virial parameter. The temperature of the scattering electrons that would produce this broadening can be estimated with $T = m_e \Delta v^2 / 16 k_B \rm{ln}(2)$ \citep{miller1991multidirectional}. We find the corresponding electron temperatures to be between $\sim10^4-10^5$ K.

It is possible that electron scattering could cause thermal broadening of this level in almost all objects in our sample. However, if we remove this estimated broadening from our values of FWHM we find the resulting widths to be systematically smaller when compared to the broad line widths in the RM sample, which are not affected by thermal broadening. It is unlikely for the maser sample to have inherently smaller values of FWHM compared to the RM objects since they live in similar host galaxies. Additionally, any geometrical effects leading to line narrowing should have similar magnitudes across both samples. Therefore, although there may be some thermal broadening in our sample, it is likely to be less than $\Delta v \approx 3000$ km s$^{-1}$.

We consider the effects more moderate thermal broadening would have on our conclusions about the likely value of $f$. To do so, we recreate the results described in Section \ref{sec:corr} after subtracting 2000 km s$^{-1}$ of thermal broadening from the FWHM of each maser object. Although this represents less thermal broadening than in NGC 1068, or the value required to match $f$ = 1.4, it allows for objects with a value of FWHM less than 2000 km s$^{-1}$ to be included in our test. Additionally, this amount of thermal broadening does not cause the maser FWHM values to be systematically smaller than the RM sample values. After subtracting this thermal broadening, we reproduce the K-S test to determine if our objects could correspond to a correlation between virial product and dynamical mass. These results are shown as the dashed lines in Figure \ref{fig:ks}. If 2000 km s$^{-1}$ thermal broadening were present in each maser object we can no longer completely reject the range of $f$ values given by \cite{woo2015black}, although we still find the value of $f = 1.4$ given by \cite{onken2004supermassive} to be unlikely. Therefore, it is possible that thermal broadening is the source of disagreement between the dynamical masses and our observed virial products. However, the true value of thermal broadening in each individual object is not known and would require more detailed observations to determine.

\subsection{Implications for the structure of the broad line region and virial masses}
\label{sec:disc_2}

There are many reasons why the single-epoch mass estimate may break down, which have been discussed thoroughly in \cite{shen2013mass}. First, this method relies upon the assumption that the BLR is virialized. In a number of AGN, measurements of different line widths and time lags using RM data show the expected virial scaling \citep[e.g.][]{peterson1999keplerian, peterson2000evidence, onken2002mass, kollatschny2003accretion}. This is not sufficient to confirm the region is virialized, however. For example, radiation pressure would lead to a similar scaling \citep{krolik2001systematic}. Even if the BLR were not virialized, the measured line widths would not be expected to deviate significantly from the expected virial value \citep{shen2013mass}.

Another issue may be the difference in measurement between the radius and width used in the virial product. \cite{krolik2001systematic} considers the possibility that the weighting over radial distribution used to determine line width could be different than that for radius of the BLR. Therefore, the product of these values would not be an accurate estimate of the enclosed mass.

The value used for width is another area of concern, as either the FWHM or $\sigma_{\rm{line}}$ could be used. FWHM is used more commonly because it can be measured more easily than $\sigma_{\rm{line}}$, which often requires modeling \citep{dalla2020sloan}. Additionally, measurements of $\sigma_{\rm{line}}$ can vary depending on the choice of method \citep[e.g.][]{denney2009systematic, rafiee2011biases, rafiee2011supermassive, assef2011black}, leading to different values of BH mass \citep[e.g.][]{shen2013mass}. However, use of FWHM may introduce bias into the mass measurement \citep[e.g.][]{rafiee2011biases, dalla2020sloan}.

The assumption of a single $f$ value is not necessarily valid either. The standard $f$ value is calculated with the M-$\sigma_*$ relation, which is assumed to hold between AGN and quiescent galaxies in the calibration of $f$. However, within the dynamical sample there is some evidence for different scaling relations for different galaxy morphologies \citep[e.g.][]{hu2008black, greene2008black, graham2008populating, graham2009m, hu2009black, gultekin2009m, greene2010precise, mcconnell2013revisiting}. At this point, it is unclear whether the maser galaxies behave differently from spirals without nuclear activity \citep{greene2016megamaser}. Therefore, the choice of objects can produce different values of $f$ \citep[e.g.][]{shen2013mass}.

We find $f$ to vary between our observed galaxies, but must consider the limitations of our method for determining $f$ for megamaser galaxies. Due to the orientation of these objects, we can only measure the broad lines in polarized light. As discussed in Section \ref{sec:linewidth}, this could introduce unaccounted for bias into our calculation. However, when we add the RM sample we see very similar results, suggesting that maser measurements alone are not biased.

\section{Summary and Future Work}
\label{sec:summary}

We have used spectropolarimetric measurements of megamaser galaxies with known dynamical BH masses to determine the accuracy of the single-epoch method. We do not find strong evidence for a correlation between the virial product and dynamical mass. Additionally, $f$ was found to vary significantly between objects and was not found to correlate with any observable parameters. Although we cannot rule out a correlation between virial product and dynamical mass, we show that this correlation is unlikely for specific values of $f$ previously proposed in the literature. We supplement our sample with RM-modeled objects, and find consistent results.

Further observations would be necessary to calibrate the virial product and reach a better determination of the value of $f$. Additional spectropolarimetric measurements of megamaser galaxies may also provide stronger evidence for a lack of correlation between virial product and dynamical mass. Multi-object RM happening now with SDSS, and planned to continue with SDSS-V, will yield new information about BLR \citep{shen2014sloan, homayouni2020sloan}. Additionally, Las Cumbres \citep{brown2013cumbres}, and in the future Rubin Observatory \citep{bianco2021optimization, abell2009lsst}, will provide high-cadence monitoring of AGN that will hopefully produce new insight into the BLR structure. ELTs will in principle measure dynamical BH masses to $z>1$ \citep{gultekin2019astro2020}, further expanding the possible comparison sample.

Our understanding of the cosmic evolution of BH mass density and BH-galaxy scaling relations often relies on single-epoch virial masses \citep[e.g.][]{laor1998quasar, wandel1999central, mclure2002measuring, vestergaard2006determining, kelly2013demographics, volonteri2016inferences, pensabene2020alma}. If geometry and other unknown factors significantly affect $f$ or the broad line width, then the weak relation between black hole mass and virial product may introduce large unquantified uncertainties in the inference of a mass. Such uncertainties make individual measurements of BHs very challenging, and these measurements should be approached cautiously, particularly in small samples of objects.

\section*{Acknowledgments}

We thank Michael DiPompeo for providing the observations included in this paper and Yue Shen for useful discussions. We also thank the referee for a timely and constructive report. All of the observations reported in this paper were obtained with the Southern African Large Telescope (SALT) under program 2017-1-SCI-002 (PI: DiPompeo). J.E.G. and A.D.G. acknowledge support from the National Science Foundation grant AAG/1007094. R.C.H. acknowledges support from the National Science Foundation through CAREER award number 1554584.

\software{PySALT \citep{crawford2010pysalt}, Astropy \citep{astropy:2013, astropy:2018}, WebPlotDigitizer \citep{Rohatgi2020}}

\bibliography{references}{}
\bibliographystyle{aasjournal}

\appendix

\section{Spectropolarimetry}
\label{sec:app_a}

We show the spectropolarimetry data for all remaining sample objects.

\begin{figure*}[htb!]
	\plotone{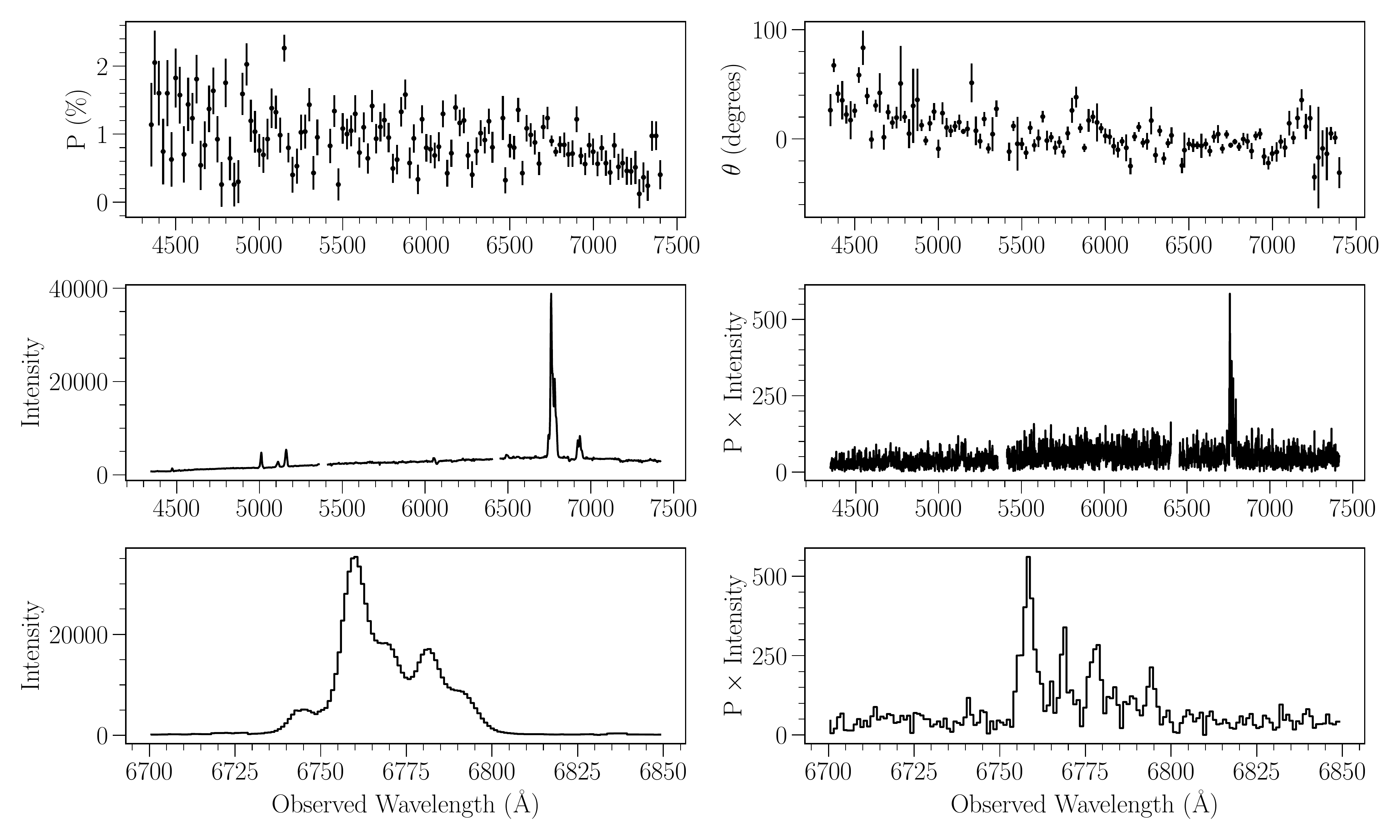}
    \caption{Same as Figure \ref{fig:ic2560_spectra}, but showing MRK 1029.}
    \label{fig:mrk1029_spectra}
\end{figure*}

\begin{figure*}[htb!]
	\plotone{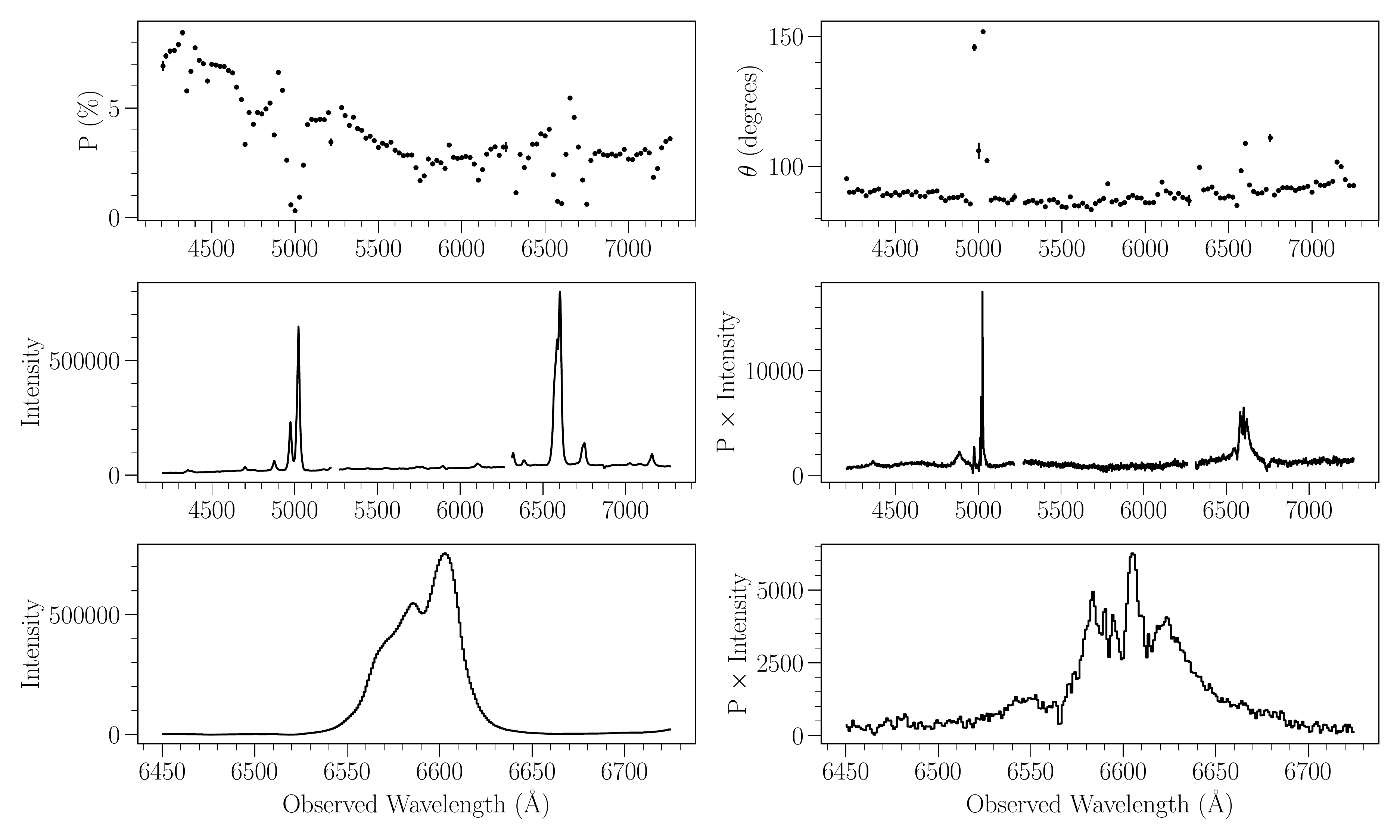}
    \caption{Same as Figure \ref{fig:ic2560_spectra}, but showing NGC 1068.}
    \label{fig:ngc1068_spectra}
\end{figure*}

\begin{figure*}[htb!]
	\plotone{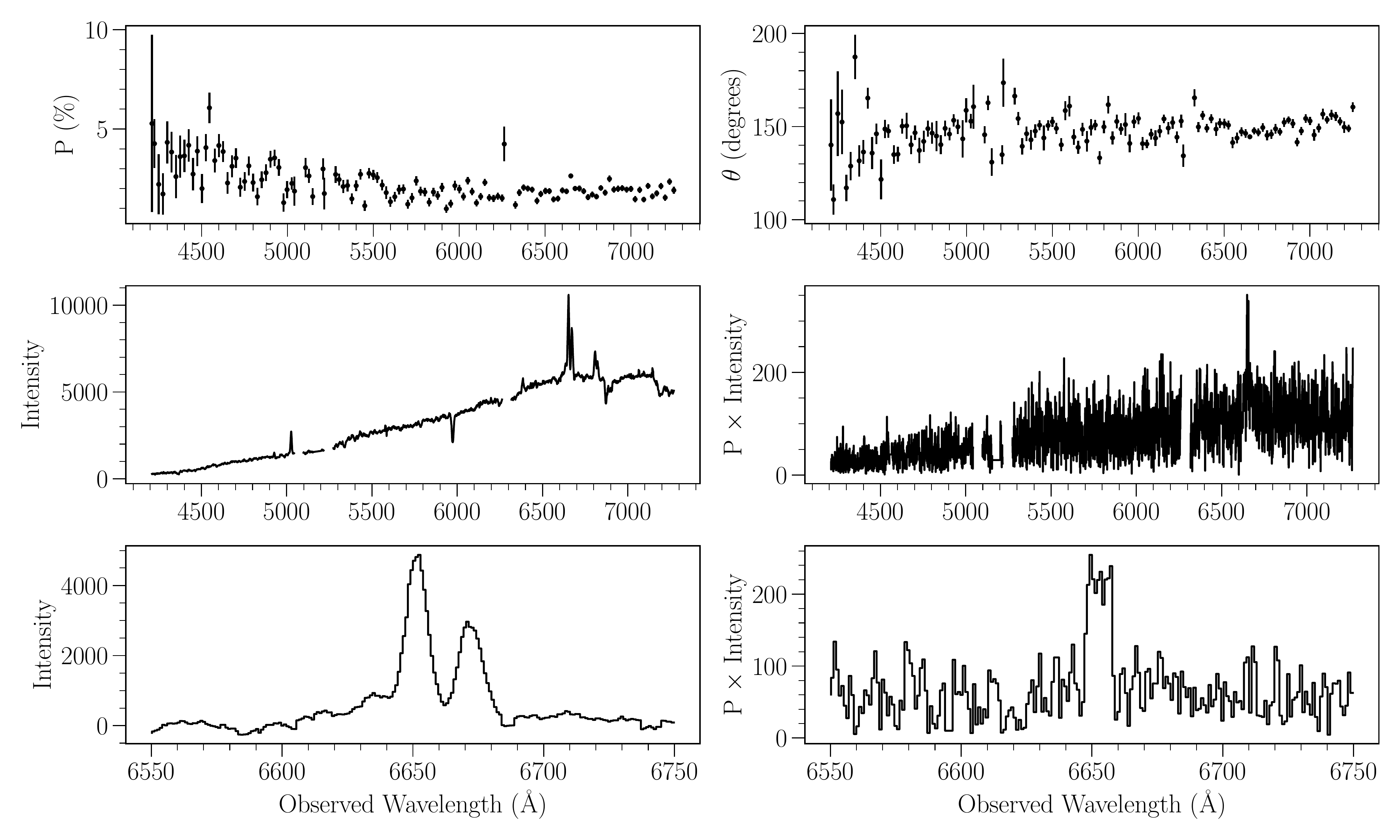}
    \caption{Same as Figure \ref{fig:ic2560_spectra}, but showing NGC 1194.}
    \label{fig:ngc1194_spectra}
\end{figure*}

\begin{figure*}[htb!]
	\plotone{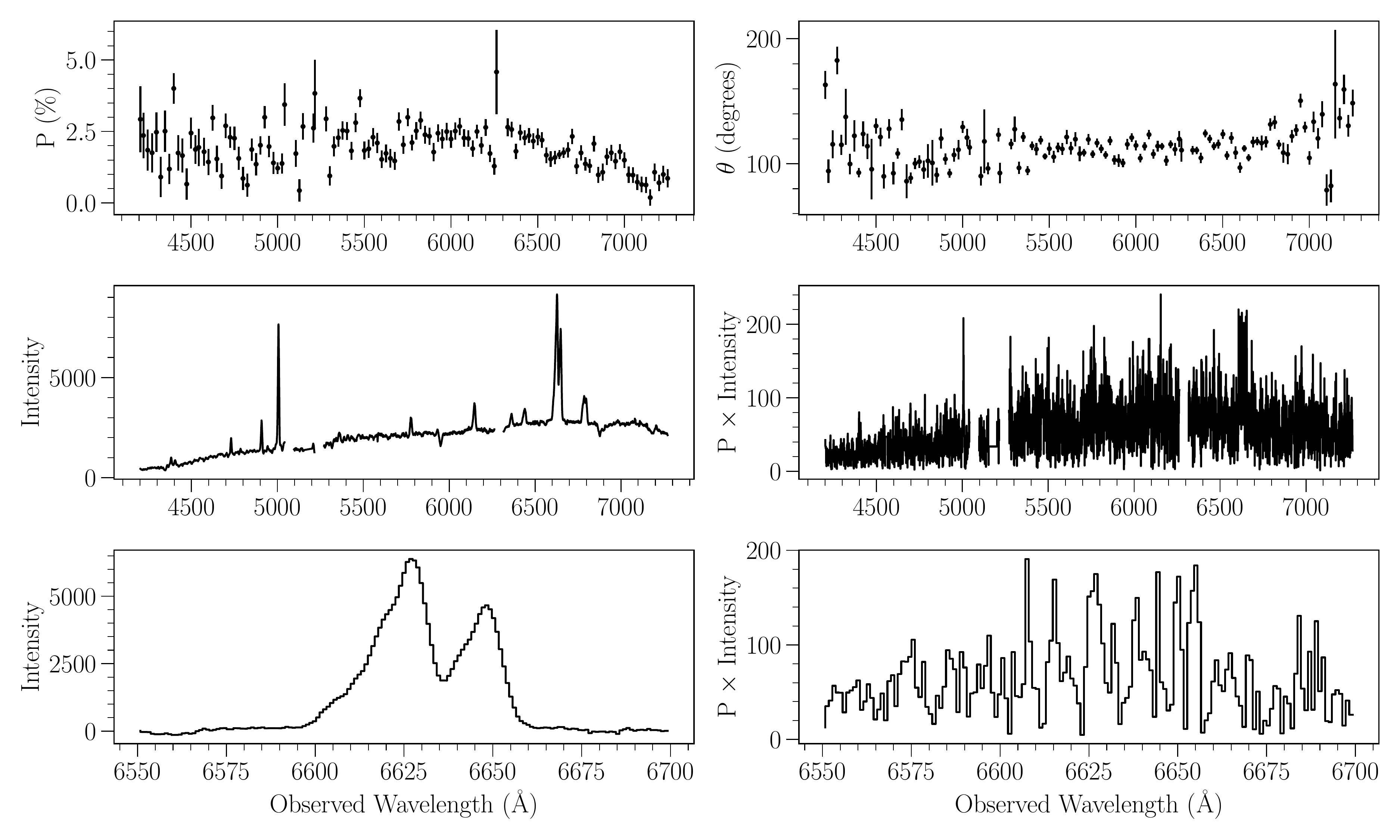}
    \caption{Same as Figure \ref{fig:ic2560_spectra}, but showing NGC 1320.}
    \label{fig:ngc1320_spectra}
\end{figure*}

\begin{figure*}[htb!]
	\plotone{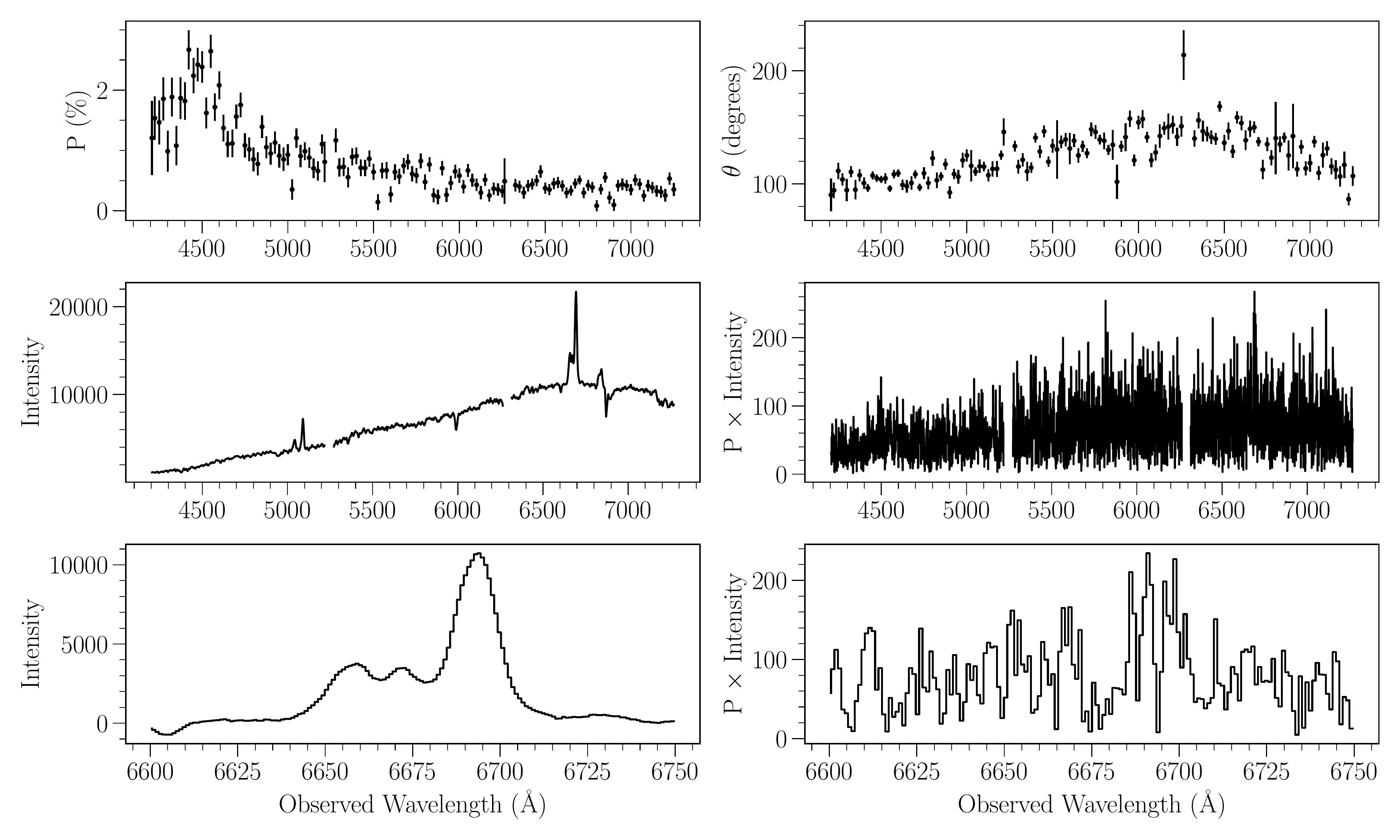}
    \caption{Same as Figure \ref{fig:ic2560_spectra}, but showing NGC 2960.}
    \label{fig:ngc2960_spectra}
\end{figure*}

\begin{figure*}[htb!]
	\plotone{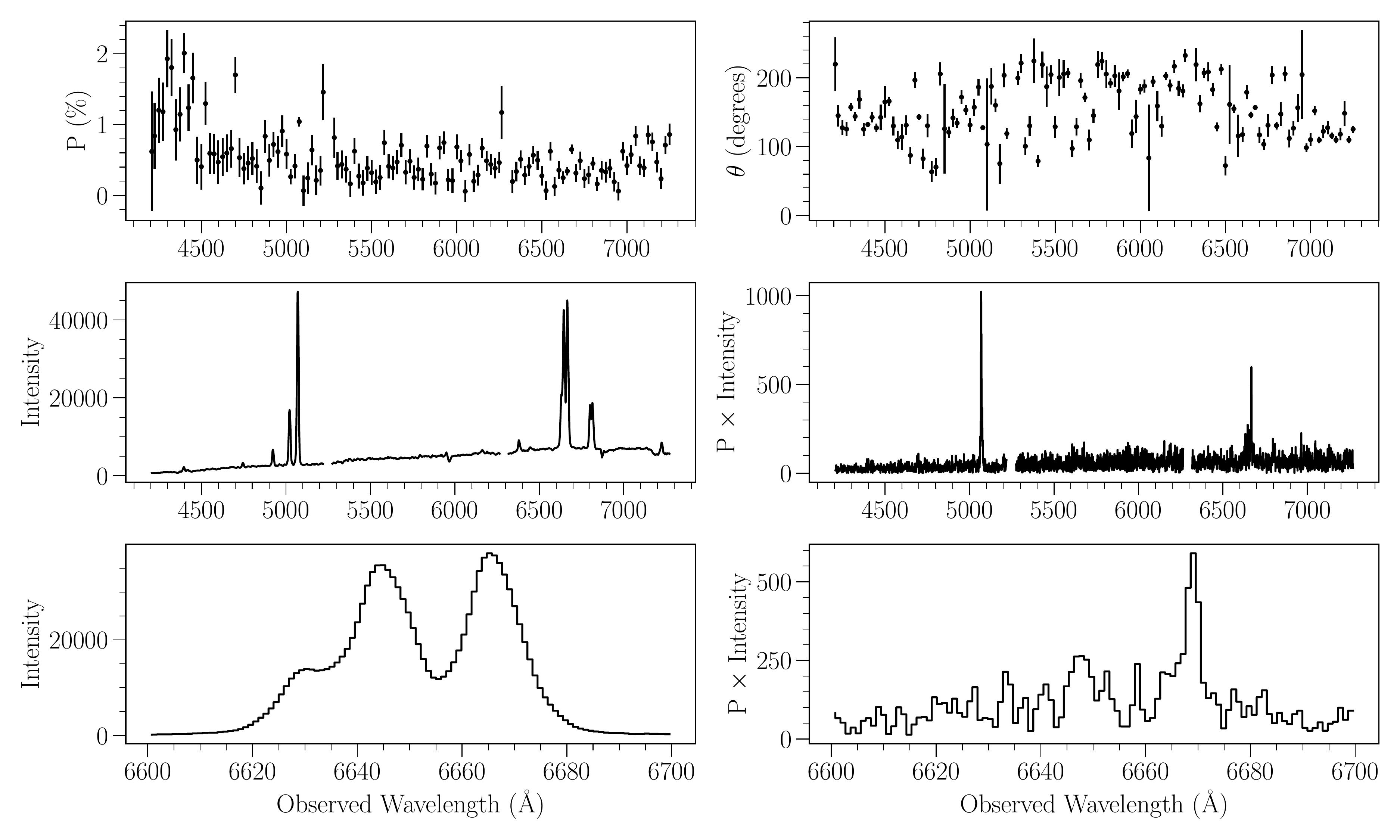}
    \caption{Same as Figure \ref{fig:ic2560_spectra}, but showing NGC 3393.}
    \label{fig:ngc3393_spectra}
\end{figure*}

\begin{figure*}[htb!]
	\plotone{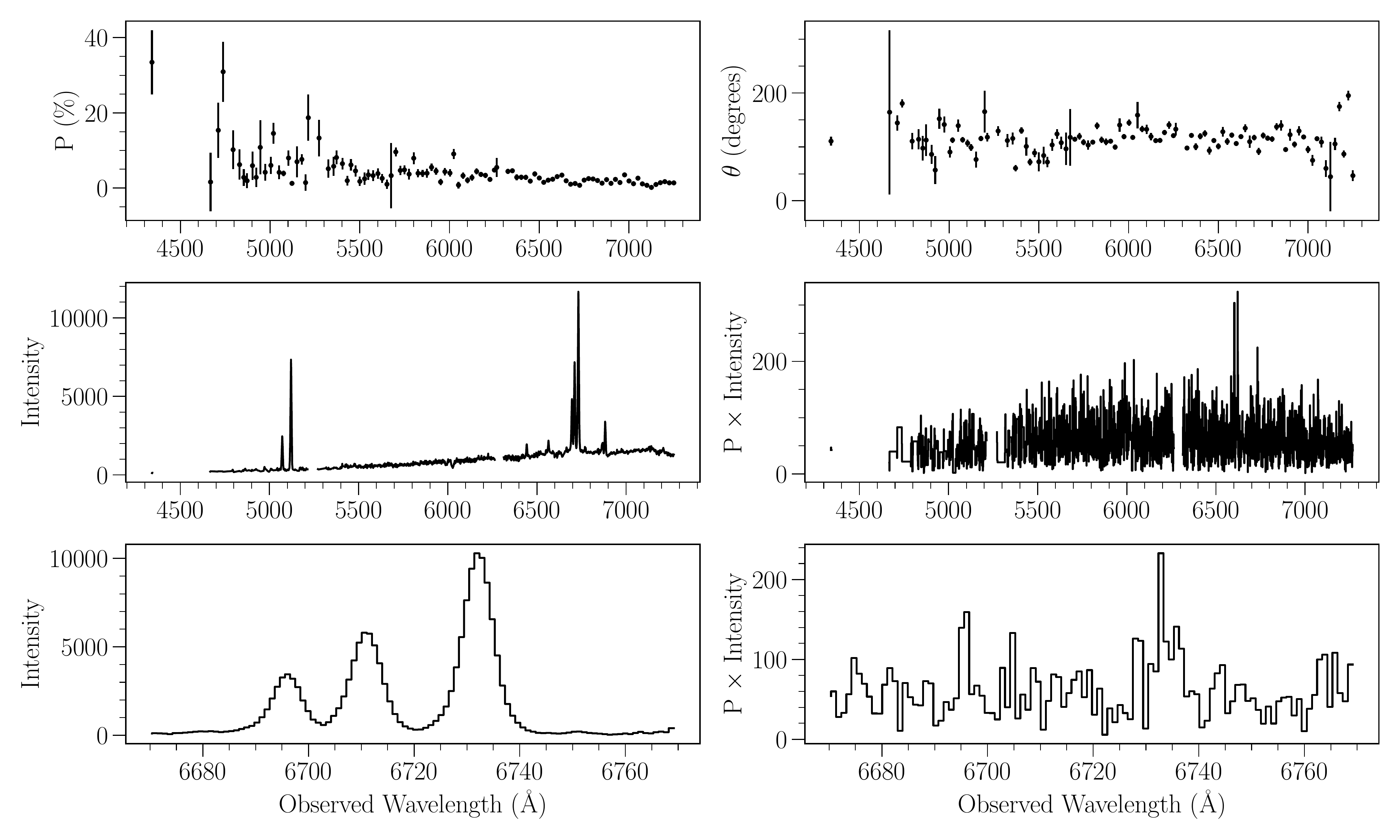}
    \caption{Same as Figure \ref{fig:ic2560_spectra}, but showing NGC 5495.}
    \label{fig:ngc5495_spectra}
\end{figure*}

\begin{figure*}[htb!]
	\plotone{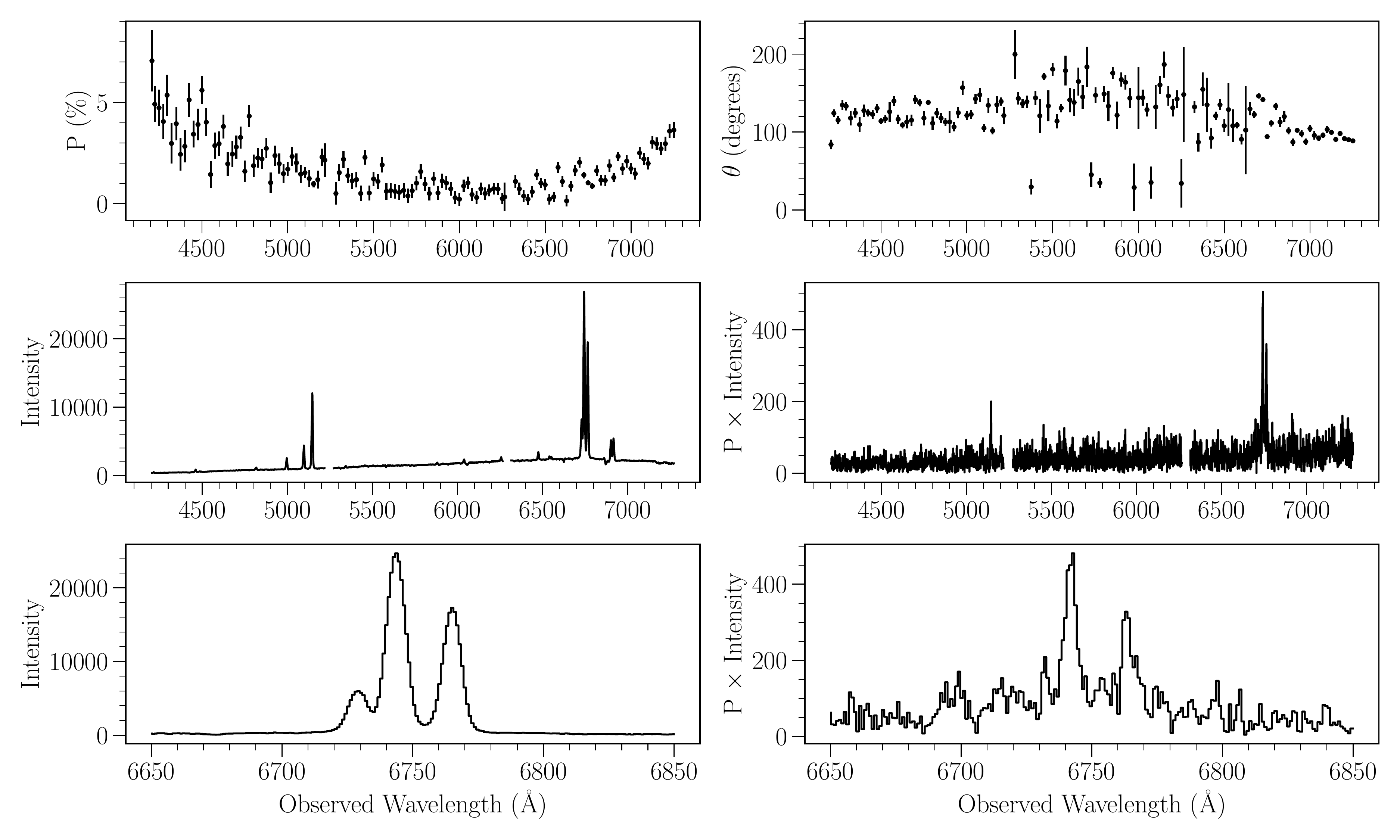}
    \caption{Same as Figure \ref{fig:ic2560_spectra}, but showing NGC 5765b.}
    \label{fig:ngc5765b_spectra}
\end{figure*}

\section{Correlation Tests}
\label{sec:app_b}

Table \ref{tab:app_corr} lists the results of all correlation tests described in Section \ref{sec:corr}. This includes correlations between additional variable pairs, and correlations for the maser sample excluding possible outliers.

\begin{deluxetable}{llcc}
\tablecaption{\label{tab:app_corr} Complete Correlation Test Results}
\tablewidth{0pt}
\renewcommand{\arraystretch}{1.3}
\tablehead{
\colhead{Objects} & \colhead{Comparison} & \colhead{r} & \colhead{p}
}
\decimalcolnumbers
\startdata
        All Masers & $\rm{M_{BH}}$ - log$_{10}$f & $0.69_{-0.13, -0.29}^{+0.10, +0.19}$ & $0.04_{-0.03, -0.04}^{+0.08, +0.24}$\\
& $\rm{M_{BH}}$ - Virial Product & $-0.05_{-0.20, -0.34}^{+0.24, +0.49}$ & $0.68_{-0.27, -0.48}^{+0.22, +0.30}$\\
& $\rm{M_{BH}}$ - FWHM & $-0.04_{-0.18, -0.32}^{+0.20, +0.41}$ & $0.73_{-0.24, -0.46}^{+0.19, +0.25}$\\
& $\rm{M_{BH}}$ - $\rm{L_x}$ & $-0.22_{-0.10, -0.18}^{+0.13, +0.29}$ & $0.57_{-0.17, -0.29}^{+0.23, +0.39}$\\
& $\rm{M_{BH}}$ - R & $-0.19_{-0.19, -0.35}^{+0.23, +0.49}$ & $0.57_{-0.28, -0.44}^{+0.29, +0.41}$\\
& $\rm{M_{BH}}$ - log$_{10}(\sigma_*$) & $0.36_{-0.13, -0.27}^{+0.13, +0.23}$ & $0.35_{-0.16, -0.25}^{+0.22, +0.48}$\\
& log$_{10}$f - $\rm{L_x}$ & $-0.42_{-0.15, -0.29}^{+0.20, +0.45}$ & $0.26_{-0.15, -0.23}^{+0.30, +0.65}$\\
& log$_{10}$f - FWHM & $-0.52_{-0.13, -0.24}^{+0.17, +0.40}$ & $0.15_{-0.09, -0.13}^{+0.20, +0.58}$\\
& log$_{10}$f - $\lambda_{\rm{edd}}$ & $-0.29_{-0.19, -0.36}^{+0.23, +0.50}$ & $0.43_{-0.25, -0.38}^{+0.34, +0.53}$\\
& log$_{10}$f - log$_{10}(\sigma_*$) & $-0.04_{-0.18, -0.36}^{+0.18, +0.36}$ & $0.75_{-0.26, -0.49}^{+0.17, +0.24}$\\
                \hline
        No NGC 4258 & $\rm{M_{BH}}$ - log$_{10}$f & $0.57_{-0.25, -0.53}^{+0.19, +0.31}$ & $0.14_{-0.11, -0.13}^{+0.30, +0.72}$\\
& $\rm{M_{BH}}$ - Virial Product & $0.15_{-0.24, -0.42}^{+0.29, +0.58}$ & $0.64_{-0.37, -0.60}^{+0.25, +0.34}$\\
& $\rm{M_{BH}}$ - FWHM & $0.22_{-0.20, -0.38}^{+0.23, +0.42}$ & $0.60_{-0.32, -0.51}^{+0.28, +0.38}$\\
& log$_{10}$f - $\rm{L_x}$ & $-0.12_{-0.34, -0.58}^{+0.39, +0.72}$ & $0.52_{-0.33, -0.48}^{+0.33, +0.46}$\\
& log$_{10}$f - FWHM & $-0.37_{-0.23, -0.39}^{+0.30, +0.66}$ & $0.36_{-0.24, -0.33}^{+0.38, +0.60}$\\
& log$_{10}$f - $\lambda_{\rm{edd}}$ & $-0.17_{-0.32, -0.57}^{+0.37, +0.72}$ & $0.52_{-0.34, -0.49}^{+0.33, +0.45}$\\
                \hline
        No NGC 4258, 5765b & $\rm{M_{BH}}$ - log$_{10}$f & $0.22_{-0.38, -0.73}^{+0.33, +0.56}$ & $0.52_{-0.34, -0.48}^{+0.33, +0.46}$\\
& $\rm{M_{BH}}$ - Virial Product & $0.50_{-0.42, -0.79}^{+0.30, +0.42}$ & $0.25_{-0.22, -0.25}^{+0.47, +0.71}$\\
& $\rm{M_{BH}}$ - FWHM & $0.47_{-0.30, -0.63}^{+0.22, +0.37}$ & $0.29_{-0.20, -0.27}^{+0.38, +0.66}$\\
& log$_{10}$f - $\rm{L_x}$ & $-0.06_{-0.41, -0.69}^{+0.46, +0.82}$ & $0.50_{-0.34, -0.48}^{+0.33, +0.47}$\\
& log$_{10}$f - FWHM & $-0.47_{-0.25, -0.40}^{+0.35, +0.83}$ & $0.27_{-0.20, -0.26}^{+0.40, +0.68}$\\
& log$_{10}$f - $\lambda_{\rm{edd}}$ & $-0.04_{-0.41, -0.71}^{+0.44, +0.78}$ & $0.53_{-0.35, -0.50}^{+0.32, +0.45}$\\
                \hline
        Masers and RM Modeling & $\rm{M_{BH}}$ - log$_{10}$f & $0.46_{-0.11, -0.23}^{+0.10, +0.18}$ & $0.02_{-0.02, -0.02}^{+0.06, +0.25}$\\
& $\rm{M_{BH}}$ - FWHM & $0.06_{-0.14, -0.27}^{+0.16, +0.31}$ & $0.61_{-0.34, -0.55}^{+0.27, +0.37}$\\
& $\rm{M_{BH}}$ - log$_{10}(\sigma_*$) & $0.40_{-0.12, -0.25}^{+0.11, +0.21}$ & $0.07_{-0.05, -0.06}^{+0.15, +0.45}$\\
& log$_{10}$f - L & $0.07_{-0.08, -0.15}^{+0.08, +0.15}$ & $0.72_{-0.24, -0.44}^{+0.20, +0.27}$\\
& log$_{10}$f - FWHM & $-0.36_{-0.08, -0.14}^{+0.08, +0.16}$ & $0.08_{-0.05, -0.07}^{+0.10, +0.27}$\\
& log$_{10}$f - $\lambda_{\rm{edd}}$ & $-0.17_{-0.12, -0.22}^{+0.14, +0.28}$ & $0.40_{-0.25, -0.35}^{+0.37, +0.56}$\\
& log$_{10}$f - log$_{10}(\sigma_*$) & $-0.12_{-0.11, -0.22}^{+0.12, +0.23}$ & $0.57_{-0.28, -0.46}^{+0.29, +0.41}$\\
        \hline
\enddata
\tablecomments{Full results of Pearson's $r$ test described in Section \ref{sec:corr}. The $r$ and $p$ values shown are the median along with bounds containing 68 and 95\% of the random samples.}
\end{deluxetable}

\section{Object Information}
\label{sec:app_c}

Table \ref{tab:app_big} compiles information for all megamaser galaxies included in Table \ref{tab:fwhm} and objects with RM modeling from \cite{pancoast2014modelling, grier2017structure}, and \cite{williams2018lick}.

\begin{deluxetable*}{lccccc}
\tablecaption{\label{tab:app_big} Maser and RM Sample Information}
\tablewidth{0pt}
\renewcommand{\arraystretch}{1.2}
\tablehead{
\colhead{Object} & \colhead{log(M$_{\rm{BH}}$/M$_\odot$)} & \colhead{Ref.} & \colhead{log L$_{5100}$ (erg/s)} & \colhead{log($\sigma_*$) (km s$^{-1}$)} & \colhead{Ref.}
}
\decimalcolnumbers
\startdata
Circinus & 6.06 $\pm$ 0.1 & 1 & 42.6 & 1.9 $\pm$ 0.02 & 1\\
IC 2560 & 6.64 $\pm$ 0.06 & 1 & 43.8 & 2.15 $\pm$ 0.03 & 1\\
Mrk 1210 & 7.15 $\pm$ 0.006 & 2 & 43.7 & 1.91 $\pm$ 0.08 & 6\\
NGC 1068 & 6.92 $\pm$ 0.25 & 1 & 43.8 & 2.18 $\pm$ 0.02 & 1\\
NGC 2273 & 6.88 $\pm$ 0.02 & 2 & 43.4 & 2.1 $\pm$ 0.03 & 1\\
NGC 3393 & 7.20 $\pm$ 0.33 & 1 & 43.7 & 2.17 $\pm$ 0.03 & 1\\
NGC 4258 & 7.58 $\pm$ 0.03 & 1 & 41.6 & 2.06 $\pm$ 0.04 & 1\\
NGC 4388 & 6.92 $\pm$ 0.01 & 2 & 43.0 & 2.0 $\pm$ 0.04 & 1\\
NGC 5765b & 7.66 $\pm$ 0.04 & 2 & 43.4 & 2.21 $\pm$ 0.05 & 1\\
Arp 151 & 6.62$_{-0.13}^{+0.1}$ & 3 & 42.5 & 2.07 $\pm$ 0.02 & 7\\
Mrk 1310 & 7.42$_{-0.27}^{+0.26}$ & 3 & 42.2 & 1.92 $\pm$ 0.03 & 7\\
NGC 5548 & 7.51$_{-0.14}^{+0.23}$ & 3 & 43.0 & 2.29 $\pm$ 0.03 & 7\\
NGC 6814 & 6.42$_{-0.18}^{+0.24}$ & 3 & 42.0 & 1.98 $\pm$ 0.01 & 7\\
SBS 1116+583A & 6.99$_{-0.25}^{+0.32}$ & 3 & 42.1 & 1.96 $\pm$ 0.02 & 7\\
Mrk 335 & 7.25 $\pm$ 0.1 & 4 & 43.7 & 1.8 $\pm$ 0.1 & 8\\
Mrk 1501 & 7.86$_{-0.17}^{+0.2}$ & 4 & 44.3 & 2.3 $\pm$ 0.1 & 9\\
3C 120 & 7.84$_{-0.19}^{+0.14}$ & 4 & 43.9 & 2.21 $\pm$ 0.05 & 10\\
PG 2130+099 & 6.92$_{-0.23}^{+0.24}$ & 4 & 44.1 & 2.21 $\pm$ 0.05 & 11\\
Mrk 50 & 7.50$_{-0.18}^{+0.25}$ & 5 & 42.9 & 2.04 $\pm$ 0.06 & 12\\
Mrk 141 & 7.46$_{-0.21}^{+0.15}$ & 5 & 43.4 & 2.38 $\pm$ 0.02 & 13\\
Mrk 279 & 7.58 $\pm$ 0.08 & 5 & 43.0 & 2.29 $\pm$ 0.03 & 14\\
Mrk 1511 & 7.11$_{-0.17}^{+0.2}$ & 5 & 43.1\\
NGC 4593 & 6.65$_{-0.15}^{+0.27}$ & 5 & 42.4 & 2.13 $\pm$ 0.02 & 14\\
PG 1310-108 & 6.48$_{-0.18}^{+0.21}$ & 5 & 43.4\\
Zw 229-015 & 6.94 $\pm$ 0.14 & 5 & 42.7\\
\enddata
\tablecomments{Values used for correlation tests for both the maser and RM sample. Columns 2-3: Dynamical or modeled mass and reference. Maser dynamical masses are taken from sources listed in each reference. Column 4: Optical luminosity from RM modeling or calculated as described in Section \ref{sec:masscalc}. Columns 5-6: Stellar dispersion for all megamaser and RM modeled objects included in Section \ref{sec:BHmass}. Values are taken from sources listed in references. References: (1) \cite{greene2016megamaser}, (2) \cite{kuo2020megamaser}, (3) \cite{pancoast2014modelling}, (4) \cite{grier2017structure}, (5) \cite{williams2018lick}, (6) \cite{marinucci2012link}, (7) \cite{woo2010lick}, (8) \cite{botte2005stellar}, (9) \cite{dasyra2007host}, (10) \cite{nelson1995stellar}, (11) \cite{grier2013stellar}, (12) \cite{barth2011lick}, (13) \cite{greene2006measuring}, (14) \cite{nelson2004relationship}}
\end{deluxetable*}

\end{document}